\documentclass[twocolumn,showpacs,preprintnumbers,amsmath,amssymb,aps]{revtex4}
\usepackage{epsfig}
\usepackage{dcolumn}
\usepackage{bm}
\DeclareGraphicsExtensions{.eps,.ps,.eps.gz,.ps.gz,.eps.Z,{}}

\newcommand{\beq}{\begin{equation}} 
\newcommand{\eeq}{\end{equation}} 
\newcommand{\beqa}{\begin{eqnarray}} 
\newcommand{\eeqa}{\end{eqnarray}} 
\newcommand{\bea}{\begin{array}} 
\newcommand{\ea}{\end{array}} 

\newcommand{\dd}{{\rm d}}
\renewcommand{\pl}{\partial}
\newcommand{\inta}{\int_{-i\infty}^{+i\infty}} 
\newcommand{\lag}{\langle} 
\newcommand{\rag}{\rangle}
\newcommand{\law}{\stackrel{\rm law}{=}}
\newcommand{\ii}{{\rm i}}

\newcommand{\Om}{\Omega_{\rm m}}
\newcommand{\cH}{{\cal H}}
 
\newcommand{\bv}{{\bf v}}
\newcommand{\bu}{{\bf u}}
\newcommand{\bs}{{\bf s}}
\newcommand{\bx}{{\bf x}}
\newcommand{\bq}{{\bf q}}
\newcommand{\thetat}{\tilde{\theta}}
\newcommand{\ut}{\tilde{u}}
\newcommand{\psit}{\tilde{\psi}}
\newcommand{\cP}{{\cal P}}

\newcommand{\cD}{{\cal D}}
\newcommand{\Rpsi}{R^{\psi}}
\newcommand{\Rpsit}{\tilde{R}^{\psi}}
\newcommand{\Rrho}{R^{\delta}}
\newcommand{\deltat}{\tilde{\delta}}
\newcommand{\Rrhot}{\tilde{R}^{\delta}}
\newcommand{\cJ}{{\cal J}}
\newcommand{\Ai}{\mbox{Ai}}
\newcommand{\Aip}{\mbox{Ai}\,'}
\newcommand{\om}{\omega}
\newcommand{\kappat}{\tilde{\kappa}}
\newcommand{\Rkap}{R^{\kappa}}
\newcommand{\Rkappsi}{R^{\kappa,\psi}}
\newcommand{\Rkapt}{\tilde{R}^{\kappa}}
\newcommand{\Rkappsit}{\tilde{R}^{\kappa,\psi}}
\newcommand{\Rkapb}{\bar{R}^{\kappa}}

\begin{document}

\title{Eulerian and Lagrangian propagators for the adhesion model (Burgers dynamics)}

\author{Francis Bernardeau}
\affiliation{Service de Physique Th{\'e}orique,
         CEA/DSM/SPhT, Unit{\'e} de recherche associ{\'e}e au CNRS, CEA/Saclay,
         91191 Gif-sur-Yvette c{\'e}dex}
\author{Patrick Valageas}
\affiliation{Service de Physique Th{\'e}orique,
         CEA/DSM/SPhT, Unit{\'e} de recherche associ{\'e}e au CNRS, CEA/Saclay,
         91191 Gif-sur-Yvette c{\'e}dex}
\vspace{.2 cm}

\date{\today}
\vspace{.2 cm}

\begin{abstract}
Motivated by theoretical studies of gravitational clustering in the Universe, we compute
propagators (response functions) in the adhesion model. This model,
which is able to reproduce the skeleton of the cosmic web and includes nonlinear
effects in both Eulerian and Lagrangian frameworks, also corresponds to the Burgers
equation of hydrodynamics. Focusing on the one-dimensional case with power-law
initial conditions, we obtain exact results for Eulerian and Lagrangian propagators.
We find that Eulerian propagators can be expressed in terms of the one-point velocity
probability distribution and show a strong decay at late times and high wavenumbers, interpreted as a
``sweeping effect'' but not a genuine damping of small-scale structures.
By contrast, Lagrangian propagators can be written in terms of the shock mass function
-- which would correspond to the halo mass function in cosmology -- and saturate to
a constant value at late times. Moreover, they show a power-law dependence on scale or
wavenumber which depends on the initial power-spectrum index and is directly related
to the low-mass tail of the shock mass function.  
These results strongly suggest that Lagrangian propagators are much more sensitive probes
of nonlinear structures in the underlying density field and of relaxation processes
than their Eulerian counterparts.
\end{abstract}

 \vskip2pc

\maketitle

\section{Introduction}
\label{sec:intro}

The formation of the large-scale structures observed in the present Universe
is an important topic of modern cosmology \cite{Peebles1980}. In the standard
scenario, these large-scale structures (such as galaxies, clusters of galaxies,
filaments and large voids) have formed through the amplification by gravitational
instability of small primordial Gaussian fluctuations, with a nearly scale-invariant
power spectrum, generated by an early inflationary stage. 
Within the concordant model of cosmology
(see for instance \cite{2003ApJS..148..175S} for comparison with current observational data), where the dynamics has been governed since $z \sim 3000$
by a collisionless dark matter component (although dark energy dominates the expansion at about $z<1$), 
this gives rise to a hierarchical evolution, as increasingly larger
scales turn nonlinear (very large scales being in the linear Gaussian regime and small
scales in the highly nonlinear regime). Today, scales beyond $\sim 10$ Mpc are
well described by linear theory while scales below $\sim 1$ Mpc are within the highly
nonlinear regime.

The highly nonlinear regime has proved very difficult to handle by analytical tools
so far (see for instance \cite{2002PhR...367....1B}), and one must resort to numerical simulations and phenomenological models
(such as the halo model \cite{2002PhR...372....1C}) which involve some free parameters
that are fitted to numerical results.
However, weakly nonlinear scales remain within the reach of
systematic analytical methods, based on various perturbative schemes.
This regime is of great practical interest, as it is the focus of several observational
probes that aim at measuring the recent expansion history of the Universe
and the growth of density fluctuations to constrain cosmology. This is the case
for instance of baryon acoustic oscillation studies \cite{Eisenstein1998,Eisenstein2005}
and weak lensing surveys \cite{Munshi2008}.
Moreover, on these large scales the system can be described through an hydrodynamical
approach \cite{Peebles1980}, which greatly simplifies the problem as compared with
the full Vlasov equation that would be required to describe small scales where
multi-streaming plays a key role \cite{Valageas2004}.

This has led to renewed interest in perturbation theory techniques, in order to increase
the range where analytical results apply. In the standard perturbative approach (see 
\cite{2002PhR...367....1B} for a review), the equations of motion for the density contrast and velocity
fields,  $\delta(\bx,t)$ and $\bv(\bx,t)$, are solved as expansions over powers of the
linear growing modes $\delta_L(\bx,t)$ and $\bv_L(\bx,t)$. Then, truncating these
series at a given order and next performing the average over the Gaussian initial
conditions yields the statistical quantities of interest, such as the density power spectrum
(which is the Fourier transform of the two-point density correlation).
As described in \cite{Crocce2006a,Crocce2006b}, it is possible to reorganize these
perturbative expansions in terms of statistical quantities of interest, such as the
two-point correlations and propagators, and to perform partial resummations to improve
the convergence. Alternatively, writing the equations of motion in terms of the
two-point correlations and propagators themselves, one obtains an infinite hierarchy of
equations (because one starts from nonlinear equations of motion) that must be truncated
at some order. This can be done in many ways, using systematic expansions of
path-integral formulations of the dynamics
\cite{Valageas2007a,Valageas2008,Matarrese2007} or closure methods that
truncate the hierarchy, written as a set of relations between the $n$ and $n+1$
correlations, by using some approximation at a level $n+1$ to obtain a closed system
of equations for lower-order quantities \cite{Taruya2008,Pietroni2008}.
These methods can also be applied within a Lagrangian framework
\cite{Matsubara2008,BernardeauVal2008}, where the dynamics is written in terms
of the displacement field.
A simultaneous comparison of such approaches with numerical simulations is presented
in \cite{Carlson2009}.

Most of these alternative approaches to the standard perturbative expansion
involve the so-called propagators, $G(\bx,t;\bq)$, that can be seen as the cross-correlation
between a nonlinear field $\phi(\bx,t)$ and its initial condition $\phi_0(\bq)$
(or equivalently its linear growing mode $\phi_L(\bq,t)$), as
$G(\bx,t;\bq)=\lag \phi(\bx,t) \phi_0(\bq) \rag$. They are also restrictions to $t_2=0$ of the
more general response functions $R(\bx_1,t_1;\bx_2,t_2)$, that measure the mean
change $\Delta\phi(\bx_1,t_1)$ of the field at time $t_1$ generated by an infinitesimal
perturbation $\Delta\phi(\bx_2,t_2)$ at the earlier time $t_2<t_1$.
The partial resummation proposed in \cite{Crocce2006a,Crocce2006b} predicts
a Gaussian decays in the nonlinear regime, of the form $e^{-D_+^2k^2}$
where $D_+(t)$ is the linear density growth rate, which is in good agreement with
numerical simulations, whereas some other approaches only yield a power-law
decay \cite{Valageas2007a,Taruya2008}. However, as explained in
\cite{Valageas2007b} this strong damping is due to a ``sweeping effect'' rather than
to a genuine loss of memory associated with relaxation processes. Thus, extending
such resummation schemes to a Lagrangian framework \cite{BernardeauVal2008},
one finds that Lagrangian propagators keep growing in the nonlinear regime instead
of decaying (within these approximations).

In order to shed light on the behavior of these response functions it is desirable
to obtain their properties for some closely related dynamics where exact
results can be derived. This would help understanding which physical processes
govern their behavior and building more realistic approximations. The simplest
dynamics that arises in this context is the Zel'dovich approximation \cite{1970A&A.....5...84Z},
where the particle displacement field is set equal to its linear prediction. Then,
explicit expressions can be obtained for the correlation and response functions
\cite{Crocce2006a,Valageas2007b}, and different-time Eulerian statistics show a Gaussian
decay of the form $e^{-D_+^2 k^2}$ at high wavenumbers, whereas Lagrangian
statistics remain equal by construction to their linear prediction.
This simplified dynamics is useful to show how such a high-$k$ decay is produced.
However, because its Lagrangian structure is so simple
and it is grossly inaccurate in the highly nonlinear regime (particles keep escaping
to infinity after shell-crossing) it is insufficient to shed light on the strongly nonlinear
regime and on the generic properties of both Eulerian and Lagrangian statistics.
 
A second dynamics, which is able to handle some non-linear effects in both Eulerian
and Lagrangian frameworks, is provided by the adhesion model \cite{Gurbatov1989}.
This adds to the Zel'dovich dynamics an infinitesimal viscosity so that particles
cannot cross.
This binds collapsed structures and fixes (at least at a qualitative level)
the main failure of the Zel'dovich approximation (but the dynamics of collapsed halos
remains non-trivial, see \cite{BernardeauVal2009b}).
Numerical simulations show that this
simple dynamics already provides a significant improvement over the Zel'dovich dynamics
and is able to reproduce the large-scale skeleton of the cosmic web \cite{Weinberg1990}.
On the other hand, as seen in \cite{Gurbatov1989}, the adhesion model also corresponds
to the standard Burgers equation \cite{Burgersbook}, which was originally introduced as a
simplified model for hydrodynamical turbulence and has been the subject of many
studies, see \cite{Bec2007} for a recent review. The advantage of this nonlinear
dynamics is that it can be explicitly integrated, and exact results can be derived in some
cases, mostly in one dimension for power-law initial conditions.

Therefore, in this article we investigate the properties of the Eulerian and Lagrangian
propagators, or response functions, $R(x,t;q_0)$, obtained within this adhesion model
(we do not consider here equal-time power spectra or correlation functions, which
have been studied in previous works, e.g. \cite{Valageas2009a,Valageas2009c}).
Since we are interested in exact analytical results, we focus on the one-dimensional
case for power-law initial conditions, where asymptotic tails can be derived in both
linear and highly nonlinear regimes. We also consider in more details the two
representative cases of Brownian and white-noise initial velocity, where explicit
expressions can be obtained for the full propagators.
We first recall in section~\ref{sec:Initial} how the Zel'dovich and Burgers dynamics
can be derived from the cosmological gravitational dynamics and we present the
one-dimensional power-law initial conditions that we consider in this paper. 
We also recall how the Burgers dynamics can be integrated through the Hopf-Cole
transformation \cite{Cole1951,Hopf1950} and its geometrical interpretation,
and we describe the self-similar evolution that is obtained for these power-law initial
conditions. Then, we study the Eulerian propagators in section~\ref{Eulerian-propagators},
and we recover the ``sweeping effect'' described above. Next, we consider the
Lagrangian propagators in section~\ref{Lagrangian-propagators}. We derive their
relation with the density field (through the shock mass function) and find as expected
that their properties are quite different from their Eulerian counterparts.
Finally, we conclude in section~\ref{Conclusion}.

\section{Initial conditions and geometrical solution}
\label{sec:Initial}

\subsection{The adhesion model}
\label{The-adhesion-model}

We briefly recall here how the Burgers equation appears in the cosmological
context, through the ``adhesion model'' \cite{Gurbatov1989,Vergassola1994}.
On scales much larger than the Jeans length, both the cold dark matter 
and the baryons can be described as a pressureless dust. Then,
neglecting orbit crossings one can use a hydrodynamical description governed 
by the equations of motion \cite{Peebles1980},
\beqa
\frac{\pl\delta}{\pl\tau} + \nabla.[(1+\delta) \bv] & = & 0,
\label{continuity1} \\
\frac{\pl\bv}{\pl\tau} + \cH \bv + (\bv .\nabla) \bv & = & - \nabla \phi,
\label{Euler1} \\
\Delta \phi & = & \frac{3}{2} \Om \cH^2 \delta , 
\label{Poisson1}
\eeqa
where $\tau=\int \dd t/a$ is the conformal time (and $a$ the scale factor), 
$\cH=\dd\ln a/\dd\tau$ the conformal expansion rate, and $\Om$ the matter
density cosmological parameter.
Here, $\delta$ is the matter density contrast and $\bv$ the peculiar velocity.
Since the vorticity field decays within linear theory \cite{Peebles1980},
the velocity is usually taken to be a potential field, so that $\bv$ is fully specified
by its divergence, $-\theta$, or by its potential, $\psi$, with
\beq
\theta= -\nabla.\bv , \;\;\; \bv= -\nabla \psi , \;\;\; \mbox{whence} \;\;\; 
\theta=\Delta \psi .
\label{thetachi}
\eeq
In the linear regime, one finds that the linear growing mode satisfies (using a subscript
$L$ for linearized quantities)
\beq
\theta_L = f \cH \delta_L \;\;\; \mbox{whence} \;\;\;
\phi_L = \frac{3\Om\cH}{2f} \psi_L ,
\label{thetaLdeltaL}
\eeq
where $f(\tau)$ is defined from the linear growing rate $D_+(\tau)$ 
of the density contrast by $f=\dd\ln D_+/\dd\ln a$, and  
$D_+(\tau)$ is the growing solution of
\beq
\frac{\dd^2 D_+}{\dd\tau^2}+\cH\frac{\dd D_+}{\dd\tau} = 
\frac{3}{2}\Om\cH^2 D_+ .
\label{D+}
\eeq
If we make the approximation that relation (\ref{thetaLdeltaL}) remains 
valid in the nonlinear regime, that is, we replace the Poisson equation 
(\ref{Poisson1}) by the second Eq.(\ref{thetaLdeltaL}),
$\phi=3\Om\cH\psi/(2f)$, then we obtain for the Euler equation (\ref{Euler1}):
\beq
\frac{\pl\bv}{\pl\tau} + \left(1-\frac{3}{2} \frac{\Om}{f}\right) \cH \bv 
+ (\bv .\nabla) \bv = 0.
\label{Euler2}
\eeq
Obviously, as shown by Eq.(\ref{Euler2}), within this approximation the 
velocity field now evolves independently of the density field.
As is well known \cite{Gurbatov1989}, approximation (\ref{Euler2}) 
is actually identical to the Zel'dovich approximation \cite{1970A&A.....5...84Z}.
Indeed, a change of variables for the velocity field yields
\beq
\frac{\pl\bu}{\pl D_+} + (\bu .\nabla) \bu = 0 \;\;\; \mbox{with} \;\;\; 
\bv= \left(\frac{\dd D_+}{\dd\tau}\right) \bu .
\label{Euler3}
\eeq
Equation (\ref{Euler3}) is the equation of motion of free particles, 
$\dd\bu/\dd D_+=0$, hence the trajectories are given by
\beq
\bx= \bq + D_+(\tau) \bu_{L0}(\bq) , \;\;\;
\bv= \frac{\dd D_+}{\dd\tau} \, \bu_{L0}(\bq) ,
\label{xq}
\eeq
where $\bq$ is the Lagrangian coordinate and $\bs=D_+ \bu_{L0}$ is
the displacement field that is exactly given by the linear theory. 
Equation (\ref{xq}) is the usual definition of the Zel'dovich approximation (i.e.
setting $\bs=\bs_L$). Then, in order to prevent particles from escaping to
infinity after shell-crossing, and to mimic the trapping within gravitational
potential wells, one can add an infinitesimal viscosity, $\nu \Delta \bu$,
to the right-hand-side of Eq.(\ref{Euler3}), which becomes the standard
Burgers equation \cite{Burgersbook}.
This is the adhesion model proposed in \cite{Gurbatov1989} to study
the formation of large-scale structures.

Numerical simulations \cite{Weinberg1990} have shown that
this simplified model is able to reproduce the cosmic web seen in
gravitational simulations: starting with identical initial conditions it
recovers the shape and the location of filaments (but the latter are now
infinitesimally thin while halos are point-like objects).
From a theoretical point of view, the advantage of the adhesion model
and of the Zel'dovich approximation is to provide a simpler dynamics
which can be exactly solved (at least in a few non-trivial cases) while
remaining close to the gravitational dynamics. In particular, the
advective quadratic nonlinearities that appear in the equations of motion
(\ref{continuity1})-(\ref{Euler1}) are preserved. Then, both Zel'dovich and
Burgers dynamics may be used as benchmarks to test approximation schemes
devised for the gravitational dynamics \cite{Valageas2007b,Valageas2009b}.

For the Zel'dovich dynamics it is possible to obtain explicit expressions
for many quantities of cosmological interest, such as the nonlinear matter power spectrum 
\cite{Schneider1995,Taylor1996} and the propagators \cite{Crocce2006a,Valageas2007b}. They all
show a Gaussian damping at high wavenumbers due to the unbounded
random displacement of particles which erases all structures in the highly
nonlinear regime (as particles keep traveling with their initial velocity $\bu$
after shell-crossing). This motivates the study of the Burgers dynamics,
which avoids this spurious damping. Moreover, while the Zel'dovich dynamics
is ill-defined for initial conditions with a slope $-1<n<1$ (because of the strong
power at large wavenumbers the linear displacement field shows 
UV divergences), both the Burgers and the gravitational dynamics remain
well-defined (as viscous sticking or gravitational trapping regularize the
dynamics on small nonlinear scales). Since the range of cosmological
interest is $-3<n<1$ it is useful to study a model that covers this domain.

However, whereas both Zel'dovich and Burgers dynamics can be explicitly integrated, the expression obtained 
for the latter is much more complex, see Eqs.(\ref{psinu0})-(\ref{unu0}) below. As applications in 
cosmological context require the computation of ensemble averages over the initial 
conditions, obtaining explicit results when the relation between the initial conditions
and the final result is involved is all the more difficult.
It turns out that it is only possible to derive exact results for a few specific cases, in one dimension.
This is the reason why in the following we consider the one-dimensional Burgers equation, 
(\ref{Burg}), and focus on power-law initial conditions.

\subsection{Equations of motion and initial conditions}
\label{Equations-of-motion}

As explained above, we are led from Eq.(\ref{Euler3}) to the study of the
one-dimensional Burgers equation. Thus, making the change of notation
$D_+ \rightarrow t$ for simplicity,
we obtain  the standard one-dimensional Burgers equation for the velocity
field $u(x,t)$ in the limit of zero viscosity,
\beq
\frac{\pl u}{\pl t} + u \frac{\pl u}{\pl x} = \nu \frac{\pl^2 u}{\pl x^2}
\hspace{1cm} \mbox{with} \hspace{1cm} \nu \rightarrow 0^+ ,
\label{Burg}
\eeq
while the density field still obeys the usual continuity equation (see
\cite{BernardeauVal2009b} for more details),
\beq
\frac{\pl\rho}{\pl t} + \frac{\pl}{\pl x}(\rho u) = 0 , \hspace{1cm} 
\mbox{with} \hspace{1cm} \rho(x,0)= \rho_0 .
\label{continuity}
\eeq
Thus, the initial conditions are set at $t=0$ (which corresponds to $D_+=0$),
with a uniform density $\rho_0$ (which corresponds to the mean comoving
matter density) and a Gaussian random velocity field $u_0(q)$.
Introducing again the velocity divergence and potential as
\beq
u = - \frac{\pl\psi}{\pl x} , \hspace{1cm} 
\theta=-\frac{\pl u}{\pl x} = \frac{\pl^2\psi}{\pl x^2} ,
\label{theta_psi_def}
\eeq
and normalizing Fourier transforms as
\beq
\theta(x,t) = \int_{-\infty}^{\infty} \dd k \, e^{\ii kx} \, \thetat(k,t) ,
\eeq
the initial divergence, $-\theta_0$, is taken as Gaussian, homogeneous, and 
isotropic, so that it is fully described by its power spectrum $P_{\theta_0}(k)$ 
with
\beq
\lag\thetat_0\rag=0 , \;\;\; \lag\thetat_0(k_1)\thetat_0(k_2)\rag = 
\delta_D(k_1+k_2) P_{\theta_0}(k_1) ,
\label{Ptheta0def}
\eeq
where $\delta_D$ is the Dirac distribution. 
As in \cite{Valageas2009b}, but restricting ourselves to one dimension,
we consider power-law initial power spectra,
\beq
P_{\theta_0}(k) = \frac{D}{2\pi} \, k^{n+2} \;\;\; \mbox{with} \;\; -3<n<1 ,
\label{ndef}
\eeq
where $D$ is a normalization factor.
Since we have $\ut(k,t)=(\ii/k)\thetat(k,t)$, the initial energy spectrum
is a power law,
\beq
\lag\ut_0(k_1) \ut_0(k_2)\rag = \delta_D(k_1+k_2) E_0(k_1) ,
\label{E0def}
\eeq
with
\beq
E_0(k) = k^{-2} P_{\theta_0}(k) = \frac{D}{2\pi} \,  k^n .
\label{E0n}
\eeq

As can be seen from the analysis in \cite{Valageas2009b}, the index $n$
introduced in Eqs.(\ref{ndef})-(\ref{E0n}) also corresponds to the standard index
$n$ used in three-dimensional cosmology, where it is defined from the linear density
contrast $\delta_L$ as $P_{\delta_L}^{d=3}(k) \propto k^n$.
More precisely, in arbitrary
dimension $d$ one must define the initial power spectra as 
$P_{\theta_0}(k) \propto k^{n+3-d}$ and $E_0(k) \propto k^{n+1-d}$. Then,
many properties only depend on the index $n$, independently of dimension $d$,
such as the scaling laws (\ref{scalingtheta0}) and (\ref{selfsimilar}) below,
see \cite{Valageas2009b}.

Since for the standard CDM cosmology the local slope $n$ runs from $1$ at
large scales to $-3$ at small scales, the range (\ref{ndef}) covers the cases of
cosmological interest. This interval can actually be split into two distinct classes.
First, for $-1<n<1$, which corresponds to large power at high $k$ (``UV-class''),
the initial velocity field is homogeneous. Moreover, it is singular (e.g., a white noise
for $n=0$) but this ultraviolet divergence is regularized as soon as $t>0$ by the
infinitesimal viscosity \cite{Burgersbook}. Note that this is a non-perturbative effect
and that the Zel'dovich dynamics, which lacks this regularization process, is not
defined in this case. This class is the ``type B initial conditions'' studied in
\cite{She1992}: the initial velocity is the derivative of a fractional Brownian motion.

Second, for $-3<n<-1$, which corresponds to large power at
low $k$ (``IR-class''), the initial velocity field is no longer homogeneous but only
shows homogeneous increments \cite{Frisch1995} (thus the divergence, $-\theta$,
is still homogeneous). This is the ``type A initial conditions'' of \cite{She1992}: the initial
velocity is a fractional Brownian motion.
Then, to handle the infrared divergence at low $k$ one must
choose a reference point, such as the origin $x_0=0$, with $u_0(x_0)=0$,
and define the initial velocity in real space as
\beq
u_0(x) = \int_{-\infty}^{\infty}\dd k \, \left( e^{\ii k x}-e^{\ii k x_0} \right) \, \ut_0(k) , 
\;\; \mbox{for} \;\; -3<n<-1 .
\label{uxukn}
\eeq
Note that because of the nonlinear advective term in the Burgers equation 
(\ref{Burg}), the increments of the velocity field are no longer 
homogeneous for $t>0$, which also means that the divergence $\theta(x,t)$
is no longer homogeneous either.
However, at large distance from the reference point (i.e. taking the
limit $|x_0|\rightarrow \infty$ or $|x|\rightarrow \infty$), 
we can expect to recover
an homogeneous system (in terms of velocity increments and matter distribution),
see \cite{Frisch2005} for more detailed discussions.
This can be shown explicitly for the case $n=-2$,
where the initial velocity field is a Brownian motion 
\cite{Bertoin1998,Valageas2009a}. On the other hand, we may add a low-$k$ cutoff
to the initial power spectrum and restrict ourselves to finite times
and scales where the influence of the infrared cutoff is expected to vanish
for equal-time statistics. This property can also be seen at a perturbative level
or from numerical simulations in the three-dimensional gravitational case studied
in cosmology for equal-time statistical quantities \cite{Vishniac1983,Jain1996}.
This is due to the Galilean invariance of the equations of motion and also holds
for the Zel'dovich dynamics \cite{Valageas2007b}.

Normalizing the velocity potential by $\psi_0(0)=0$, that is,
\beq
\psi_0(x)= \int_0^x \dd x' \, u_0(x') ,
\eeq
and the velocity field by
$u_0(0)=0$ if $-3<n<-1$, the initial conditions obey the scaling laws
\beqa
\lambda>0 & : & \theta_0(\lambda x) \law \lambda^{-(n+3)/2} \; \theta_0(x) ,
\nonumber \\
&& u_0(\lambda x) \law \lambda^{-(n+1)/2} \; u_0(x) , \nonumber \\
&& \psi_0(\lambda x) \law \lambda^{(1-n)/2} \; \psi_0(x) ,
\label{scalingtheta0}
\eeqa
where ``$\law$'' means that both sides have the same statistical properties.

These initial conditions can also be expressed in terms of the linear density
field as follows. If we linearize the equations of motion (\ref{Burg})-(\ref{continuity})
we obtain the solution for $\nu=0$,
\beq
\theta_L(x,t) = \theta_0(x) , \;\;\; \delta_L(x,t) = t \, \theta_0(x) ,
\label{deltaL}
\eeq
where we defined the density contrast as $\delta(x,t)= (\rho(x,t)-\rho_0)/\rho_0$
and the subscript $L$ stands for the linear regime. (Note that we recover the
usual growing mode for the density contrast since as explained above $t$ stands
for the linear growing mode $D_+$ in the cosmological context.) Therefore,
we can as well define the initial conditions by the linear density contrast 
$\delta_L(x,t)$, which is Gaussian, homogeneous, and isotropic, with a power
spectrum
\beq
-3 <n <1 : \;\;\; P_{\delta_L}(k,t) = t^2 \, P_{\theta_0}(k) 
\propto t^2 \, k^{n+2} .
\label{PdeltaL}
\eeq
This is the manner in which initial conditions are usually defined in the
cosmological context.
(As explained above, in $d$ dimensions we would obtain
$P_{\delta_L}(k,t) \propto t^2 \, k^{n+3-d}$ which recovers the cosmological
notation for $d=3$.) 

In this article we consider in more details two representative cases where many
exact results can be obtained 
\cite{Burgersbook,Kida1979,She1992,Frachebourg2000,Valageas2009c,Sinai1992,Bertoin1998,Valageas2009a}.
First, the case $n=0$, associated with the UV-dominated range
$-1<n<1$, corresponds to a white-noise initial velocity field, normalized as
\beq
n=0 : \;\;\; \lag u_0(q)\rag=0 , \;\; \lag u_0(q_1) u_0(q_2)\rag =
D \, \delta_D(q_1-q_2) , 
\label{u0def_WN}
\eeq
\beq
\lag \psi_0(q)\rag=0 , \;\; \lag \psi_0(q_1) \psi_0(q_2)\rag = D \, q_1 , 
\;\;\; \mbox{for} \;\;\; 0 \leq q_1 \leq q_2 .
\label{psi0def_WN}
\eeq
Thus, the initial velocity potential is a bilateral Brownian motion that starts from 
the origin. 

Second, the case $n=-2$, associated with the IR-dominated range
$-3<n<-1$, corresponds to a Brownian initial velocity field, normalized as
\beq
n=-2 : \;\;\; \lag \theta_0(q)\rag=0 , \;\; \lag \theta_0(q_1) \theta_0(q_2)\rag =
D \, \delta_D(q_1-q_2) , 
\label{theta0def_Brown}
\eeq
\beq
\lag u_0(q)\rag=0 , \;\; \lag u_0(q_1) u_0(q_2)\rag = D \, q_1 , 
\;\;\; \mbox{for} \;\;\; 0 \leq q_1 \leq q_2 .
\label{u0def_Brown}
\eeq
Thus, it is now the initial velocity field which is a bilateral Brownian motion that
starts from the origin.

\subsection{Hopf-Cole solution and first-contact parabolas}
\label{First-contact-parabolas}

As is well known \cite{Hopf1950,Cole1951}, making the change of variable
$\psi(x,t)= 2\nu\ln\Xi(x,t)$ transforms the nonlinear Burgers equation into the
linear heat equation. This gives the explicit Hopf-Cole solution
\beq
\psi(x,t)= 2\nu \ln \int_{-\infty}^{\infty} \frac{\dd q}{\sqrt{4\pi\nu t}} \;
\exp \left[ \frac{\psi_0(q)}{2\nu}-\frac{(x-q)^2}{4\nu t} \right] .
\label{psinu}
\eeq
Then, in the inviscid limit $\nu \rightarrow 0^+$ the steepest-descent method
gives
\beqa
\nu \rightarrow 0^+ & : & \;\;\; \psi(x,t) = \max_q \left[ \psi_0(q)
- \frac{(x-q)^2}{2t} \right] ,
\label{psinu0} \\
&& \;\;\;   u(x,t) = u_0(q) = \frac{x-q(x,t)}{t} ,
\label{unu0}
\eeqa
where we introduced the Lagrangian coordinate $q(x,t)$ defined as the point
where the maximum in Eq.(\ref{psinu0}) is reached. In particular, this is the
Lagrangian coordinate (i.e. the initial location) of the particle that is located at
the Eulerian position $x$ at time $t$.
Here and in the following we note by the letter $q$ the Lagrangian coordinates,
which appear in the arguments of the initial fields at $t=0$, and by the letter $x$
the Eulerian coordinates, which appear in the arguments of the Eulerian fields
at any time $t>0$.
The Eulerian locations $x$ where there are two solutions, $q_-<q_+$, to the
maximization problem (\ref{psinu0}) correspond to shocks and all the matter
initially between $q_-$ and $q_+$ is gathered at $x$. At these points the velocity
is discontinuous while the density is infinite. 
The application $q \mapsto x(q,t)$ is usually called the Lagrangian map, and
$x \mapsto q(x,t)$ the inverse Lagrangian map (which is discontinuous at
shock locations) \cite{Bec2007}.
Outside of shocks Eq.(\ref{unu0}) clearly shows that one recovers the free-streaming
dynamics (\ref{Euler3})-(\ref{xq}) in the inviscid limit $\nu \rightarrow 0^+$.

The maximization problem (\ref{psinu0}) has a well-known geometrical solution
\cite{Burgersbook}. Indeed, let us consider the family of upward parabolas
$\cP_{x,c}(q)$ centered at $x$ and of height $c$, with a curvature radius $t$,
\beq
\cP_{x,c}(q) = \frac{(q-x)^2}{2 t} + c .
\label{paraboladef}
\eeq
Then, moving down $\cP_{x,c}(q)$ from $c=+\infty$, where the parabola is
everywhere well above the initial potential $\psi_0(q)$\footnote{This is possible
for the initial conditions (\ref{ndef}) since we have $|\psi_0(q)| \sim
q^{(1-n)/2}$, which grows more slowly than $q^2$ at large distances.},
until it touches the curve $\psi_0(q)$, the abscissa $q$ of this
first-contact point is the Lagrangian coordinate $q(x,t)$. If first-contact
occurs simultaneously at several points there is a shock at the Eulerian 
location $x$. One can build in this manner the inverse Lagrangian map
$x\mapsto q(x,t)$.

Finally, the continuity equation (\ref{continuity}) can also be integrated as follows
in one dimension (see \cite{BernardeauVal2009b} for a discussion of the more
complex case of higher dimension).
The conservation of matter implies that the density field is related to 
the inverse Lagrangian map, $x\mapsto q(x,t)$, and to the velocity
potential, $\psi(x,t)$, through the Jacobian
\beq
\rho(x,t) \dd x = \rho_0 \dd q ,
\label{Jacobian}
\eeq
whence, using Eq.(\ref{unu0}),
\beq 
\rho(x,t) = \rho_0 \, \frac{\pl q}{\pl x} = \rho_0 \, 
\left[ 1 - t \, \frac{\pl u}{\pl x} \right] =  \rho_0 \, 
\left[ 1 + t \, \frac{\pl^2 \psi}{\pl x^2} \right] ,
\label{rhoxqvpsi}
\eeq
which gives for the matter density contrast,
\beq
\delta(x,t) = t \, \theta(x,t) = t \, \frac{\pl^2 \psi}{\pl x^2} .
\label{deltapsi}
\eeq
Here we used the fact that particles do not cross each other, so that $x(q)$ and
$q(x)$ are monotonous increasing functions and there is no need to keep the
absolute value for the Jacobian $J=|\pl q/\pl x|$. In the last two equalities
in (\ref{rhoxqvpsi}) we used Eq.(\ref{unu0}) and the definition
of the velocity potential. One can easily check that (\ref{rhoxqvpsi}) is also
valid for shocks, which give rise to Dirac density peaks.
We can note from the last expression (\ref{deltapsi}) and Poisson's equation
that the velocity potential, $\psi$, is equal to the gravitational potential,
$\phi$, up to a normalization (and an additive quadratic term $\propto \rho_0 x^2/2$
associated with the mean density $\rho_0$). This property, which only holds at the
linear order in higher dimensions, is also at the basis of the Zel'dovich
approximation used in cosmology and recalled in section~\ref{The-adhesion-model},
see Eq.(\ref{thetaLdeltaL}) (see also the discussion in 
section~2.2.2 of \cite{Vergassola1994}). Thus, Eq.(\ref{deltapsi}) recovers the
well-known fact that the Zel'dovich approximation is actually exact in one dimension
before shell-crossing.

For the UV-class, $-1<n<1$, shocks are expected to be isolated and in finite number per unit length,
whereas between shocks the inverse Lagrangian map, $x\mapsto q(x,t)$, is constant,
so that the density field is made of a finite number of Dirac peaks per unit length amid
empty space. For the IR-class, $-3<n<-1$, shocks are expected to be dense so that the density field
is made of an infinite number of Dirac peaks per unit length (without any smooth
background). These results have
been explicitly proved for white-noise
\cite{AvellanedaE1995,Frachebourg2000} and Brownian \cite{Sinai1992} initial
velocity; they are only supported by phenomenological arguments and numerical simulations \cite{She1992} 
for generic values of $n$.

\subsection{Self-similar evolution}
\label{Self-similar-evolution}

For the initial conditions (\ref{ndef}), using the scaling laws (\ref{scalingtheta0})
one can see from the explicit solution (\ref{psinu0}) that the nonlinear Eulerian
fields obey the scaling laws
\beqa
\psi(x,t) & \law & t^{(1-n)/(n+3)} \; \psi \left( t^{-2/(n+3)} x,1 \right) ,
\label{scale_psi} \\
u(x,t) & \law & t^{-(n+1)/(n+3)} \; u \left( t^{-2/(n+3)} x,1 \right) , \\
q(x,t) & \law & t^{2/(n+3)} \; q \left( t^{-2/(n+3)} x,1 \right) \label{scale_q} , \\
\delta(x,t) & \law & \delta \left( t^{-2/(n+3)} x,1 \right) \label{scale_delta} .
\eeqa
We can check that the scaling (\ref{scale_delta}) agrees with the linear
mode (\ref{deltaL}). 
These scalings mean that the dynamics is self-similar: a rescaling of time is
statistically equivalent to a rescaling of distances, as
\beq
\lambda>0: \;\; t \rightarrow \lambda t, \;\;\; x \rightarrow \lambda^{2/(n+3)} x .
\label{selfsimilar}
\eeq
Thus, as in the standard cosmological scenario \cite{Peebles1980}, 
the system displays a hierarchical
evolution as increasingly larger scales turn nonlinear. More precisely, since in the
inviscid limit there is no preferred scale for the power-law initial conditions
(\ref{ndef}), the only characteristic scale at a given time $t$ is the
so-called integral scale of turbulence, $L(t)$, which is generated by the
Burgers dynamics and grows with time as in (\ref{selfsimilar}),
\beq
L(t)\propto t^{2/(n+3)} .
\label{Lt}
\eeq
It measures the typical distance between shocks,
and it separates the large-scale quasi-linear regime, where the energy spectrum
and the density power spectrum keep their initial power-law forms,
from the small-scale nonlinear regime, which is governed by shocks,
where the density power spectrum reaches a universal white-noise
behavior (i.e. $P_{\delta}(k,t)$ has a finite limit for $k\gg 1/L(t)$)
\cite{Frisch2001,Tribe2000,Valageas2009c}.
Note that the scalings (\ref{scale_psi})-(\ref{Lt}) hold for any dimension $d$,
provided we define the initial conditions by $P_{\delta_L}(k,t) \propto t^2
\, k^{n+3-d}$.

In order to express the scaling laws (\ref{scale_psi})-(\ref{scale_delta}) it is
convenient to introduce suitable dimensionless scaling variables,
\beq
Q= \frac{q}{L(t)} , \;\; X= \frac{x}{L(t)}, \;\; U=\frac{t u}{L(t)} , 
\label{QXdef}
\eeq
where $L(t)$ is the characteristic scale (\ref{Lt}), which we normalize as
\beq
L(t) = (2D t^2)^{1/(n+3)} ,
\label{Ltdef}
\eeq
where the constant $D$ was introduced in Eq.(\ref{ndef}).
Thus, equal-time probability distributions written in terms of these variables no
longer depend on time, and the scale $X=1$
is the characteristic length of the system, at any time. On large quasi-linear
scales, $X \gg 1$, density fluctuations are small and the distributions are
strongly peaked around their mean, with tails that are directly governed by
the initial conditions (but shocks cannot be neglected for $n>-2$).
On small nonlinear scales, $X \ll 1$, density fluctuations are large and probability
distributions show broad power-law regions
\cite{Valageas2009c,Valageas2009b,Valageas2009a}.
Note that for the associated power-law initial conditions in the three-dimensional
cosmological context, the gravitational dynamics also develops the same self-similar
evolution (\ref{Lt}), see \cite{Peebles1980}.

\section{Eulerian propagators}
\label{Eulerian-propagators}

\subsection{Relation with the velocity probability distribution}
\label{Eul-general-expressions}

We now consider the Eulerian propagator of the velocity potential, 
$\Rpsi(x,t;q_0)$, defined as the functional derivative of $\psi(x,t)$
with respect to the initial potential $\psi_0(q_0)$ at point $q_0$,
\beq
\Rpsi(x,t;q_0) = \lag \frac{\cD\psi(x,t)}{\cD\psi_0(q_0)} \rag .
\label{Rxdef}
\eeq
Then using the explicit Hopf-Cole solution (\ref{psinu}) to
perform the functional derivative and after the inviscid limit, $\nu\rightarrow 0^+$,
has been taken, we have
\beq
\Rpsi(x,t;q_0) =\lag \delta_D[ q(x,t) - q_0 ] \rag .
\label{Rxexp}
\eeq
Here $q(x,t)$ is again the inverse Lagrangian map introduced in (\ref{unu0}).
From its functional definition we can see that the Eulerian propagator
$\Rpsi(x,t;q_0)$ describes the sensitivity of the nonlinear potential at a given
time $t$ with respect to the initial conditions. This response function can also be
seen as a memory kernel, with a time-dependence that would give an estimate
of the time-scale beyond which initial fluctuations at a given wavelength appear to be
damped. Note that from the expression (\ref{Rxexp}) it obeys the sum rule
\beq
\int \dd q_0 \, \Rpsi(x,t;q_0) = 1 .
\label{Rxq0norm}
\eeq
Next, the average (\ref{Rxdef}) gives
\beq
\Rpsi(x,t;q_0) = p_x(q_0,t)  ,
\label{Rpsipxq}
\eeq
where $p_x(q,t)$ is the one-point probability distribution of the Lagrangian
coordinate $q(x,t)$. Since shocks form a set of zero measure, we can use
(\ref{unu0}) to write Eq.(\ref{Rpsipxq}) as
\beq
\Rpsi(x,t;q_0) = \frac{1}{t} \, p_x(u,t) , \;\;\; \mbox{with} \;\; u= \frac{x-q_0}{t} ,
\label{Rpsipxu}
\eeq
where $p_x(u,t)$ is the one-point Eulerian velocity probability distribution. Then,
for the IR-class $-3<n<-1$ where the velocity field is not homogeneous, as
discussed in section~\ref{Equations-of-motion}, the propagator $\Rpsi(x,t;q_0)$
is not homogeneous either, whereas it is homogeneous for the UV-class $-1<n<1$.
In the latter case, it can be useful to go to Fourier space, with the normalization
\beq
-1<n<1: \;\;\; \lag \frac{\cD\psit(k,t)}{\cD\psit_0(k_0)} \rag
= \delta_D(k-k_0) \, \Rpsit(k,t) ,
\label{Rhxdef}
\eeq
where the Dirac factor $\delta_D(k-k_0)$ is due to the statistical invariance 
through translations. This yields
\beq
-1<n<1: \;\; \Rpsit(k,t) = \int_{-\infty}^{\infty} \dd x \, e^{-\ii kx} \, \Rpsi(x,t;0) .
\label{Rpsih}
\eeq

Within the cosmological context, where one is mostly interested in the density
field, one rather considers the density propagator $\Rrho$ defined by
\beq
\Rrho(x,t;q_0) = \lag\frac{\cD\delta(x,t)}{\cD\delta_{L0}(q_0)}\rag ,
\label{Rrhodef}
\eeq
where $\delta(x,t)=(\rho(x,t)-\rho_0)/\rho_0$ is again the density contrast and
from Eq.(\ref{deltaL}) we defined $\delta_{L0}(q_0)$ as
\beq
\delta_{L0}(q_0) = \theta_0(q_0) \;\; \mbox{whence} \;\; 
\delta_L(x,t) = t \, \delta_{L0}(x) .
\label{deltaL0def}
\eeq
Then, from Eqs.(\ref{deltaL0def}) and (\ref{deltapsi}), we have in Fourier
space
\beq
\deltat_{L0}(k) = - k^2 \psit_0(k) \;\;\; \mbox{and} \;\;\; 
\deltat(k,t) = -t \, k^2 \psit(k,t) .
\label{rhopsi}
\eeq
Therefore, the density propagator (\ref{Rrhodef}) is equal to the velocity
potential propagator defined in Eq.(\ref{Rxdef}), multiplied by time $t$,
\beq
\Rrhot(k,t) = t \Rpsit(k,t)  \;\;\; \mbox{and} \;\;\; 
\Rrho(x,t;q_0) = t \Rpsi(x,t;q_0) .
\label{RrhoRh1}
\eeq
For the power-law initial conditions (\ref{ndef}) the first Fourier-space equality in
(\ref{RrhoRh1}) only holds for $-1<n<1$ as in (\ref{Rhxdef}).

It is interesting to note that Eqs.(\ref{Rpsipxu}) and (\ref{RrhoRh1}) show that
Eulerian propagators, or response functions, are unlikely to be effectual in probing the
nonlinear structures built in the underlying density field. As discussed in the following
sections, the relation (\ref{Rpsipxu}) implies a strong dependence on a ``sweeping effect''
associated with long wavelength modes of the velocity field. Besides, whatever the
magnitude of this effect, it is clear that the one-point velocity distribution does not
provide much pertinent information on the density field. Therefore, in more general dynamics
such as gravitational clustering, which should still behave in a similar fashion,
computing the Eulerian propagators in a perturbative manner up to high order, or exactly
in cases such as the Burgers dynamics studied here, is unlikely to shed much light
on the density field. This is a strong motivation to study Lagrangian propagators,
which are much more directly linked to the properties of the underlying density field
as we shall find out in Sect.~\ref{Lagrangian-propagators} below.

\subsection{Linear regime and IR-class}
\label{Eul-Brownian}

At large scales and early times density and velocity fluctuations are small
and the system is well described by the linearized equations of motion.
Note however that such a linear regime does not exist for the UV-class
$-1<n<1$, where the initial velocity variance $\lag u_0^2\rag$ shows a UV
divergence and shocks dominate the dynamics as soon as $t>0$.
For $-2<n<-1$ shocks also play a key role as soon
as $t>0$, and modify the naive linear predictions at a quantitative level
(i.e. numerical prefactors), but they do not change the qualitative behavior
of the distributions seen at large scales or early times (i.e. exponents
in the exponential tails), see \cite{Valageas2009b}. 

Thus, for the IR-class, or for generic initial conditions with a high-$k$ cutoff,
it is interesting to consider the linear predictions for
the Eulerian propagators (\ref{Rxdef}) and (\ref{Rrhodef}).
From Eq.(\ref{deltaL}) we have $\psi_L(x,t)= \psi_0(x)$ which would give at
zeroth-order $\Rpsi(x,t;q_0)=\delta_D(x-q_0)$. However, we can obtain the
first-order prediction from Eq.(\ref{Rpsipxu}). Indeed, for Gaussian initial
conditions the initial velocity probability distribution at position $q$ reads as
\beq
p_q(u_0) =\frac{1}{\sqrt{2\pi}\sigma_{u_0}(q)} \, 
e^{-u_0^2/(2\sigma^2_{u_0}(q))} ,
\label{p0u0}
\eeq
where we introduced the initial velocity variance
\beq
\sigma^2_{u_0}(q) = \lag u_0(q)^2 \rag .
\label{sigu0def}
\eeq
For initial conditions with an IR cutoff the initial velocity field can be
homogeneous, so that $\sigma^2_{u_0}$ and $p_q(u_0)$ do not depend on
position $q$, but for the IR-class $-3<n<-1$ this is not the case, as discussed in
section~\ref{Equations-of-motion} above Eq.(\ref{uxukn}).
Then, the linear prediction for the Eulerian propagators reads as
\beqa
\Rpsi_L(x,t;q_0) & = & \frac{1}{\sqrt{2\pi}t\sigma_{u_0}(x)} \, 
e^{-(x-q_0)^2/(2t^2\sigma^2_{u_0}(x))} , \;\; \label{RpsiL} \\
\Rrho_L(x,t;q_0) & = & \frac{1}{\sqrt{2\pi}\sigma_{u_0}(x)} \, 
e^{-(x-q_0)^2/(2t^2\sigma^2_{u_0}(x))} . \;\; \label{RrhoL}  
\eeqa
Of course, in the limit $t\rightarrow 0$ we recover $\Rpsi(x,0;q_0)=
\delta_D(x-q_0)$ and $\Rrho(x,0;q_0) =0$.

If the initial conditions show an IR cutoff, so that the system is homogeneous,
we can go to Fourier space as in Eq.(\ref{Rhxdef}). This yields
\beq
\Rpsit_L(k,t) = e^{-t^2k^2\sigma^2_{u_0}/2} , \;\;
\Rrhot_L(k,t) = t\, e^{-t^2k^2\sigma^2_{u_0}/2} .
\label{RpsitL}
\eeq
We can note that for the Zel'dovich dynamics particles exactly follow
the linear displacement field as recalled below Eq.(\ref{xq}). Then,
Eqs.(\ref{RpsiL})-(\ref{RpsitL}) are exact (for homogeneous systems)
\cite{Crocce2006a,Valageas2007b} and Eulerian propagators show the
characteristic Gaussian decay $e^{-t^2k^2}$ due to the random advection of
particles that eventually erases small-scale structures as particles escape to
infinity after shell-crossing.
For the adhesion model, that is the Burgers dynamics studied in this article,
Eqs.(\ref{RpsiL})-(\ref{RpsitL}) only hold in the linear regime.

If the initial conditions belong to the IR-class, $-3<n<-1$, without IR cutoff,
the initial velocity variance depends on position $q$ as
\beq
-3<n<-1 : \;\;\; \sigma^2_{u_0}(q) = D' \, |q|^{-(n+1)} ,
\label{sigmau0_IR}
\eeq
as seen from Eq.(\ref{scalingtheta0}), with $D'=D$ for $n=-2$.
Then, at any finite time $t$, as we go far from the reference point $x_0=0$
the nonlinear velocity distribution becomes dominated by the initial Gaussian
distribution (as nonlinear effects have only redistributed matter over the finite
scales $X \sim 1$ and $U \sim 1$),
\beqa
 \hspace{-1cm}  -3<n<-2, \;\;\; |x|\rightarrow \infty & : & \nonumber \\
&&  \hspace{-2cm}  p_x(u) \sim \frac{|x|^{(n+1)/2}}{\sqrt{2\pi D'}} 
\, e^{-u^2|x|^{n+1}/(2D')} .
\label{pxuIR}
\eeqa
For $-2<n<-1$ (in one dimension) shocks modify the numerical factor in the
exponential decay, but not the exponents that remain of the form
$\ln p_x(u) \sim -|x|^{n+1} u^2$ \cite{Valageas2009b}.
Note that the asymptotic behavior (\ref{pxuIR}) can be explicitly derived 
from the exact velocity probability distribution for
the Brownian case $n=-2$ \cite{Valageas2009a}. This can also be related
to the ``principle of permanence of large eddies'' in the generic
case \cite{Gurbatov1997}. 

Then, from expression
(\ref{Rpsipxu}) we can see that the Eulerian propagators vanish as soon as
$t>0$ in the limit $|x|\rightarrow \infty$, that is far from the reference point,
at fixed separation $|x-q_0|$. More precisely, in agreement with the
sum rule (\ref{Rxq0norm}), the Eulerian propagator spreads over an increasingly
large region so that it vanishes for any fixed distance $|x-q_0|$.
Indeed, as we go far from the reference point to avoid boundary effects, 
the local velocity variance becomes increasingly large, because of
long-wavelength modes, and this large collective velocity leads to a very
fast damping of Eulerian propagators (that becomes instantaneous in the
limit $|x|\rightarrow \infty$) as structures are transported over large distances.
This ``sweeping effect'', which agrees with the previous discussion below
Eq.(\ref{RpsitL}), is not a true loss of memory if the system is observed on
a global scale, as it does not imply that small-scale structures are erased
(thanks to the Galilean invariance of the equations of motion) since this
divergence arises from low-$k$ modes and is associated with almost
uniform random translations.

This motivates the study of Lagrangian propagators, introduced in
\cite{BernardeauVal2008} in the cosmological context, and considered
in section~\ref{Lagrangian-propagators} below, to go beyond this sweeping effect.
We can note that the sensitivity of Eulerian propagators to long-wavelength
modes of the velocity field can be directly seen from Eq.(\ref{Rpsipxu}),
which is not invariant through Galilean transformations.

\subsection{White-noise initial velocity}
\label{Eul.prop.white-noise}

\begin{figure}
\begin{center}
\epsfxsize=7 cm \epsfysize=5 cm {\epsfbox{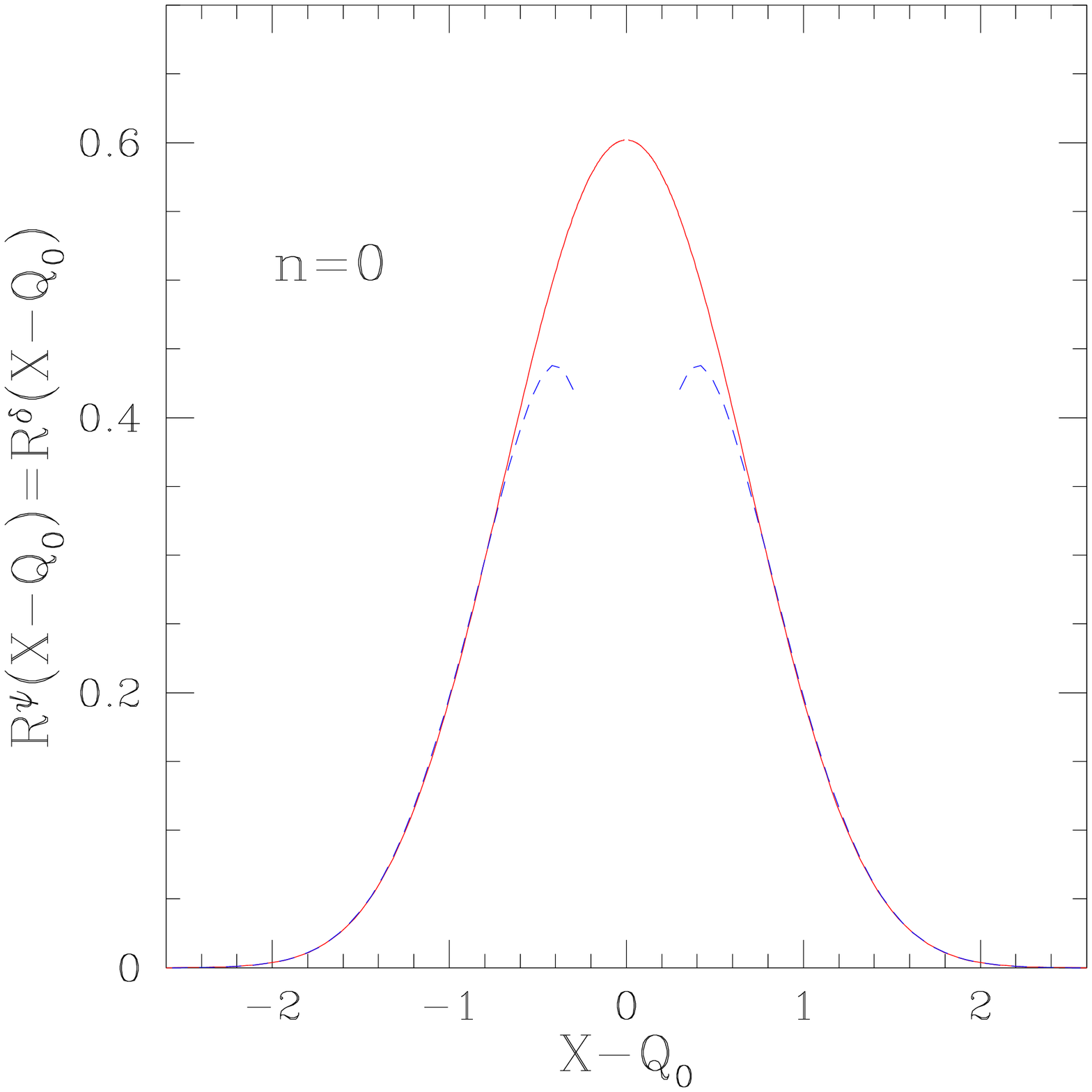}}
\epsfxsize=7 cm \epsfysize=5 cm {\epsfbox{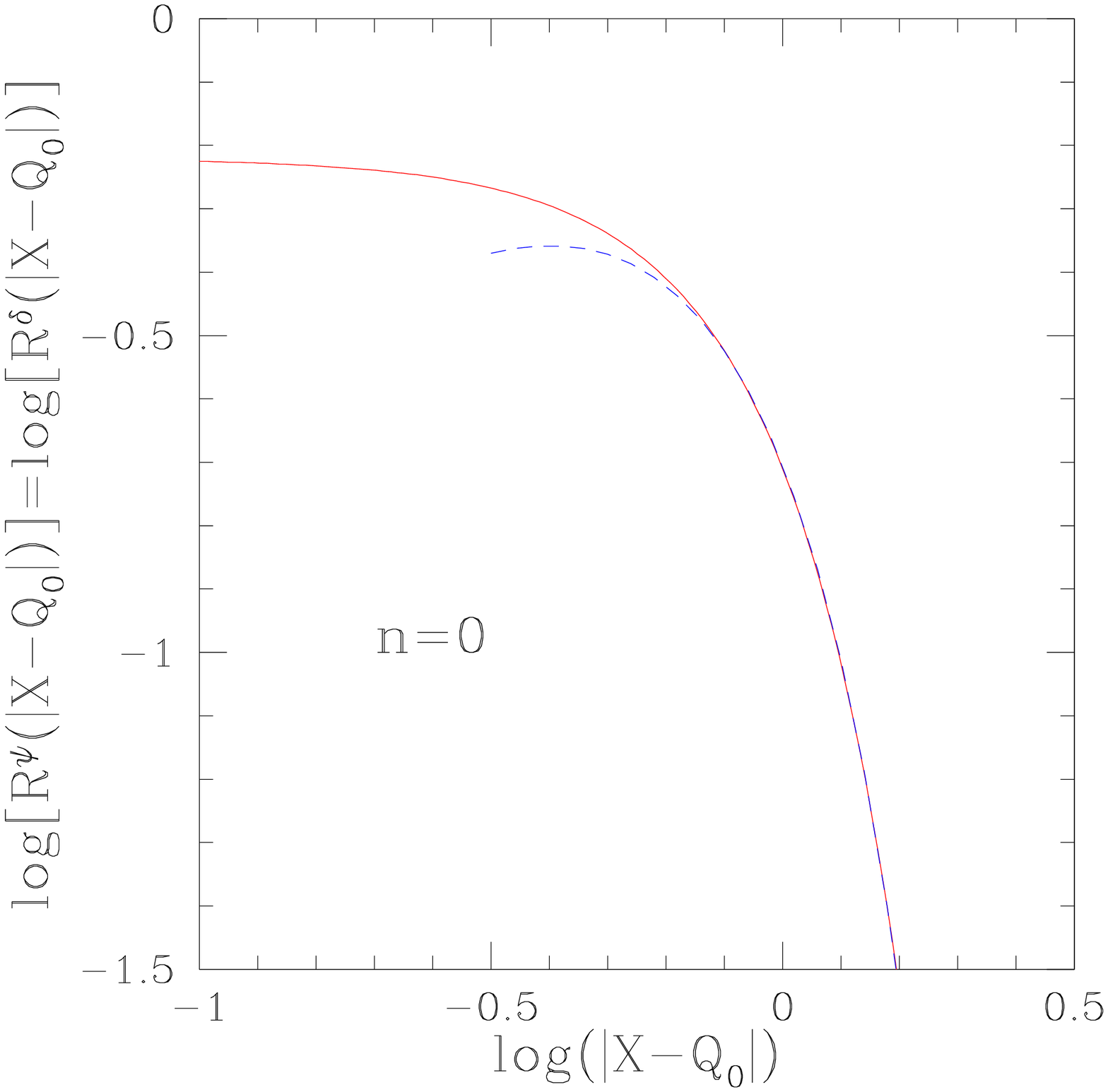}}
\end{center}
\caption{{\it Upper panel:} The Eulerian propagators
$\Rpsi(X-Q_0)=\Rrho(X-Q_0)$ obtained for $n=0$ (white-noise initial velocity),
in terms of dimensionless variables. They are also equal to the one-point velocity
distribution $P(U)$, with $U=X-Q_0$, from Eq.(\ref{RxJJ}).
The dashed lines are the asymptotic cubic exponential behavior (\ref{Rxasymp}).
{\it Lower panel:} Same as upper panel but on a logarithmic scale.}
\label{figPv}
\end{figure}

\begin{figure}
\begin{center}
\epsfxsize=7 cm \epsfysize=5 cm {\epsfbox{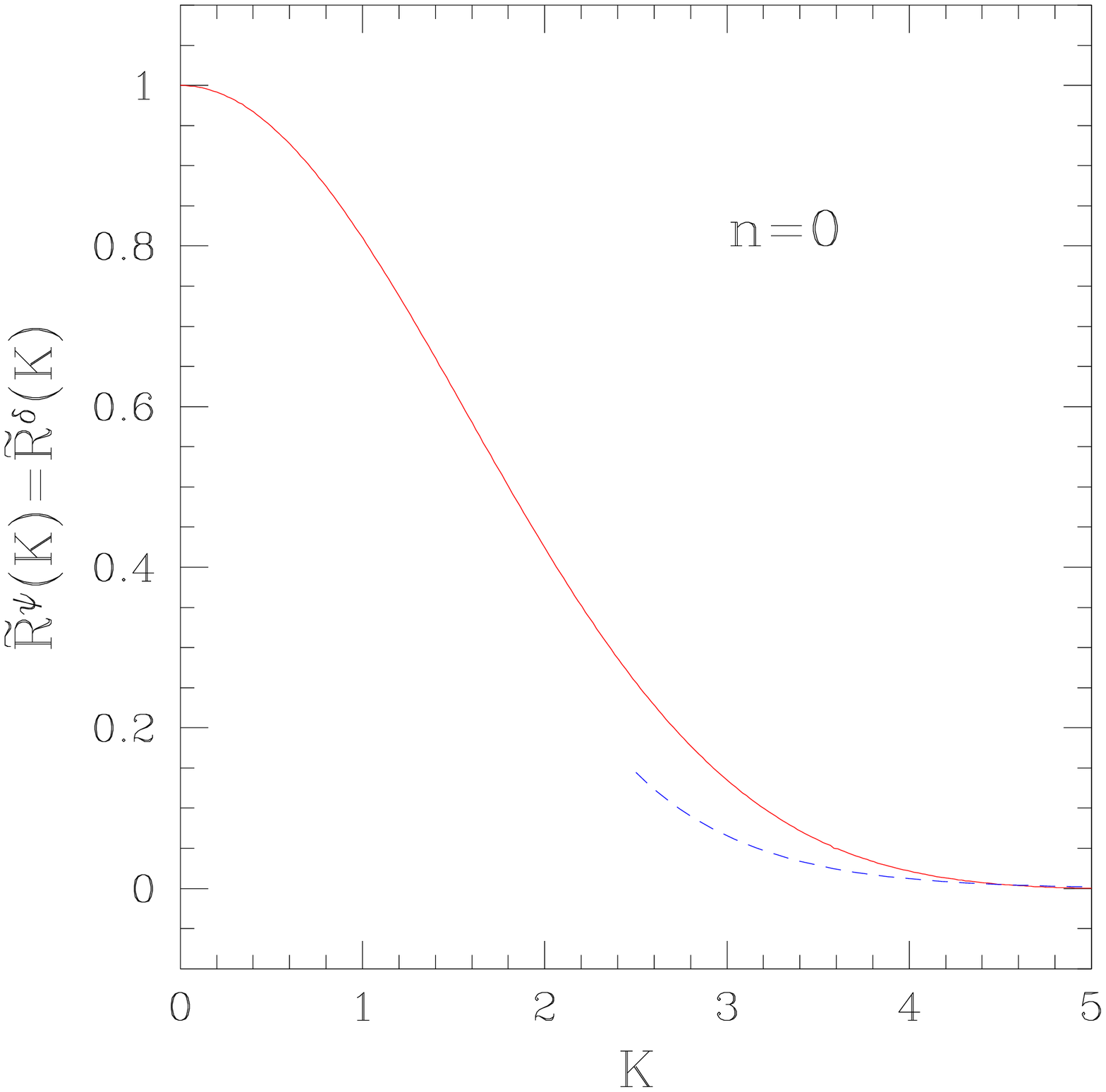}}
\epsfxsize=7 cm \epsfysize=5 cm {\epsfbox{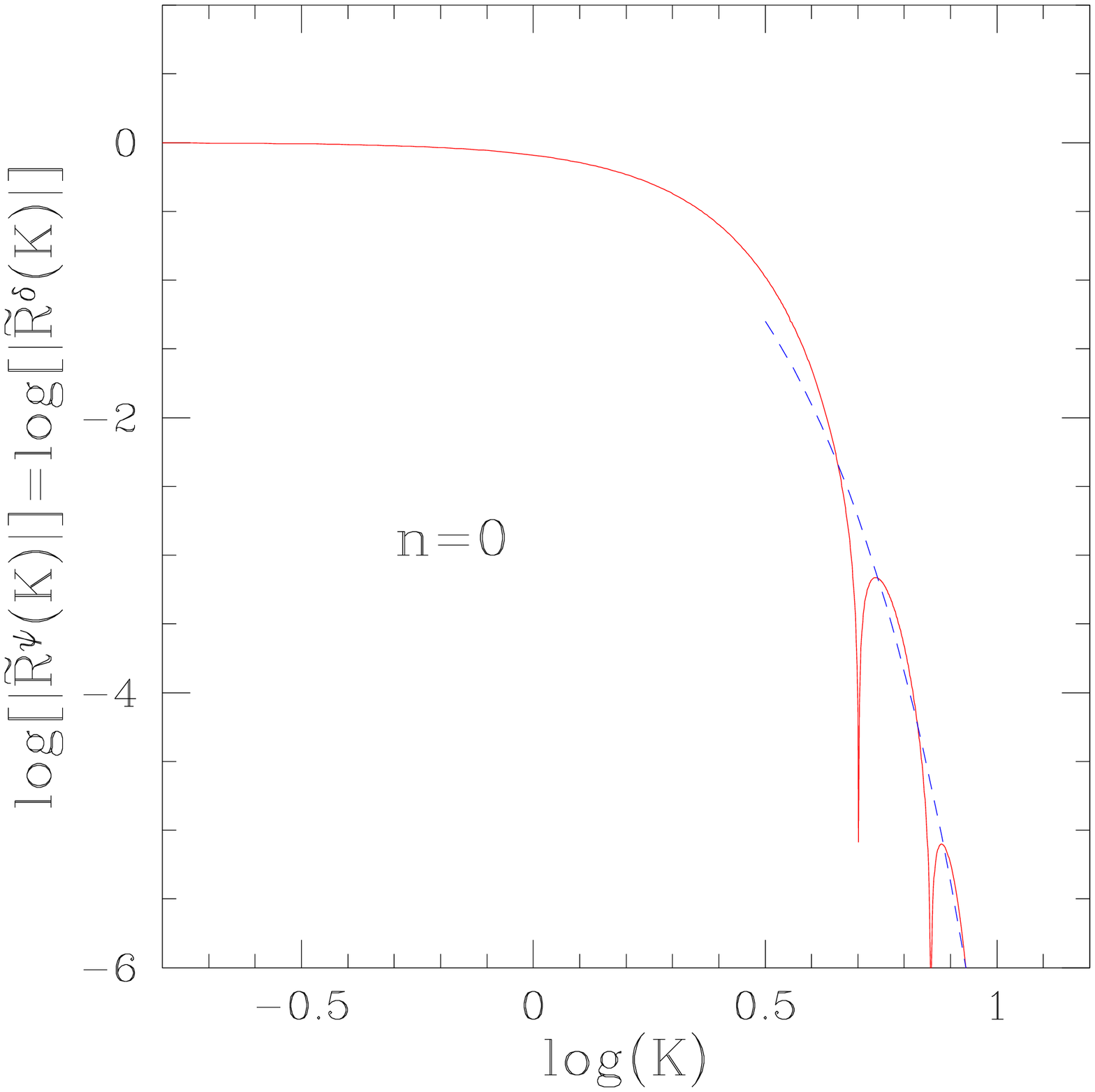}}
\end{center}
\caption{{\it Upper panel:} The Eulerian propagators $\Rpsit(K)=\Rrhot(K)$
in Fourier space, obtained for $n=0$ (white-noise initial velocity), in terms of the
dimensionless wavenumber $K$, from Eq.(\ref{RhxK}). The dashed line is the
asymptotic $3/2$-exponential behavior (\ref{RhxKasymp}) for the amplitude of $\Rpsit$,
that is setting the cosine to unity.
{\it Lower panel:} Same as upper panel but on a logarithmic scale.
The cusps correspond to changes of sign for $\Rpsit$.}
\label{figRrhok}
\end{figure}

Before turning to Lagrangian propagators, we consider the case $n=0$ of
white-noise initial velocity, that is representative of the UV-class.
For such initial conditions, $-1<n<1$, the initial variance $\sigma^2_{u_0}$
is infinite as it shows a UV divergence and the initial velocity field $u_0(q)$
is not a regular function (unless we add a UV cutoff). Then, shocks govern
the dynamics as soon as $t>0$ and the linear regime described in
Eqs.(\ref{RpsiL})-(\ref{RpsitL}) does not exist \cite{Valageas2009b}.
Thus, it is difficult to derive
generic results and we focus below on the white-noise case $n=0$
where exact results can be obtained. This should provide a qualitative
illustration of the behaviors obtained for $-1<n<1$.

In terms of the dimensionless scaling variables (\ref{QXdef}) the Eulerian
propagator (\ref{Rpsipxq}) writes as
\beq
\Rpsi(x,t;q_0) = \frac{1}{L(t)} \Rpsi(X-Q_0) ,
\label{RpsiXQ}
\eeq
with
\beq
\Rpsi(X-Q_0) = P(U) \;\;\; \mbox{and} \;\;\; U= X-Q_0 ,
\label{RxJJ}
\eeq
where $P(U)$ is the one-point probability distribution of the dimensionless
velocity $U$. The latter was obtained in \cite{Frachebourg2000} as
\beq
P(U) = \cJ(U)\cJ(-U)  \;\;\; \mbox{with} \;\;\; \cJ(U) = \inta\frac{\dd s}{2\pi\ii} \,
\frac{e^{sU}}{\Ai(s)} ,
\label{PUJJ}
\eeq
where $\Ai(x)$ is the Airy function.
Of course, because of the statistical homogeneity
and isotropy of the system, we can check that $\Rpsi(x,t;q_0)$ only depends on the
distance $|x-q_0|$ and on time, through the combination $|X-Q_0|$ thanks to the
scale-invariance (\ref{selfsimilar}).
Then, Eq.(\ref{PUJJ}) gives the asymptotic behavior at large dimensionless
separation $X$ \cite{Frachebourg2000,Valageas2009c}
\beq
|X| \gg 1 : \;\; \Rpsi(X) \sim \frac{2|X|}{\Aip(-\om_1)} \; e^{-\om_1 |X| - |X|^3/3} , 
\label{Rxasymp}
\eeq
where $-\om_1$ is the first zero of the Airy function ($\om_1\simeq 2.338$).
We display in Fig.~\ref{figPv} the Eulerian propagator $\Rpsi(X-Q_0)=P(U)$ in  terms
of dimensionless variables.

Going to Fourier space, Eqs.(\ref{Rpsih}) and (\ref{PUJJ}) yield 
\beq
\Rpsit(k,t) = \Rpsit(K) , \;\;\; \mbox{with} \;\;\; K= L(t) \, k = (2Dt^2)^{1/3} \, k ,
\eeq
where we introduced the dimensionless scaling wavenumber $K$, and
\beq
\Rpsit(K) = \inta\frac{\dd s}{2\pi\ii} \, \frac{1}{\Ai(s+\ii K/2)\Ai(s-\ii K/2)} .
\label{RhxK}
\eeq
We can check that at low wavenumber $\Rpsit(0)=1$.
Indeed, on large scales particles have not significantly moved from their initial
positions (at a finite time $t$ particles have moved over distances of order
$(2Dt^2)^{1/3}$, i.e. $X \sim 1$) so that $\Rpsi(x,t;q_0) \sim \delta_D(x-q)$ as seen
from large distances or in the limit $t \rightarrow 0$. 
More precisely, $\Rpsi(x,t;q_0)$ decays faster than any power-law over
$|x-q_0|$ at large distances, so that $\Rpsit(k,t)$ has a well-defined expansion
at $k=0$, and $\Rpsit(0,t)=1$ in agreement with the sum rule (\ref{Rxq0norm}).
Thus, at zeroth-order we recover
\beq
t \rightarrow 0 : \;\;\; \Rpsi_L(x,t;q_0) \rightarrow \delta_D(x-q_0) \;\;\;
\mbox{and} \;\;\; \Rpsit_L(k,t) \rightarrow 1 .
\label{RxL}
\eeq
This also applies to the large-scale limit $k\rightarrow 0$.
On the other hand, Eq.(\ref{RhxK}) leads to the asymptotic behavior
at large wavenumber $K$,
\beqa
K \gg 1 & : & \Rpsit(K) \sim \frac{4\sqrt{\pi}}{\Aip(-\om_1)} 
\, K^{1/4} \, e^{-\frac{\sqrt{2}}{3}K^{3/2} - \frac{\om_1}{\sqrt{2}}\sqrt{K}} \nonumber \\
&& \;\;\; \times \cos\left[ \frac{\sqrt{2}}{3}K^{3/2} - \frac{\om_1}{\sqrt{2}}\sqrt{K}
+\frac{\pi}{8} \right] .
\label{RhxKasymp}
\eeqa
Thus, the propagator $\Rpsit(K)$ decays somewhat more slowly than a Gaussian,
with increasingly fast oscillations, at high $K$.
We display the propagator $\Rpsit(K)$ in Fig.~\ref{figRrhok}. We can note that
the oscillations only appear in the very far tail where  $\Rpsit(K)$ is already
negligible.

The density propagator $\Rrho$ of Eq.(\ref{Rrhodef}) directly follows from
$\Rpsi$ using Eq.(\ref{RrhoRh1}).
In terms of dimensionless variables, this gives
\beq
\Rrho(x,t;q_0) = \frac{t}{L(t)} \, \Rrho(X-Q_0) , \;\; 
\Rrhot(k,t) = t \, \Rrhot(K) , 
\eeq
with
\beq
\Rrho(X-Q_0)= \Rpsi(X-Q_0) , \;\; \Rrhot(K)= \Rpsit(K) .
\label{RrhoRh2}
\eeq
Thus, the density propagator shows the same exponential-like decay at large 
wavenumbers or late times obtained in (\ref{RhxKasymp}) for $\Rpsit(k,t)$.

For generic index $n$ it is difficult to obtain the exact form of the velocity distribution
$p_x(u)$, which is needed to derive the Eulerian propagators.
However, as shown in \cite{Molchan1997,Ryan1998}, for $-1<n<1$ we have
the asymptotic large-velocity tails
\beq
-1<n<1 ,  \;\;\; |u| \rightarrow \infty : \;\;\; p(u) \sim e^{-t^{n+1}|u|^{n+3}} ,
\label{pu_asymp}
\eeq
where we do not specify prefactors and numerical factors in the exponents.
This gives
\beq
|x-q_0|  \rightarrow \infty : \;\;\; \Rpsi(x,t;q_0) \sim e^{-|x-q_0|^{n+3}/t^2} .
\label{Rpsi_asymp}
\eeq
As discussed in \cite{Valageas2009b} these behaviors are related to rare-event
distributions. By contrast,
the high-wavenumber tail in Fourier space depends on the details of the
small-scale highly nonlinear processes and is more difficult to estimate.
Note that the large-scale decay (\ref{Rpsi_asymp}) depends on $n$, contrary
to the Gaussian decay obtained in Eqs.(\ref{RpsiL})-(\ref{RrhoL}) for the IR-class.

As is clear from the expression (\ref{RxJJ}) and as discussed in
section~\ref{Eul-Brownian}, the Eulerian propagators
$\Rpsi(x,t;q_0)$ and $\Rrho(x,t;q_0)$ are governed by the one-point velocity 
distribution. We can note that various approximation methods, based on resummation
schemes or field-theoretic methods, have been recently devised to estimate
such propagators in the case of the gravitational or Zel'dovich dynamics
\cite{Crocce2006a,Matarrese2007,Valageas2007a,Valageas2007b}, and they show
a Gaussian or power-law decay that is set by the variance of the initial velocity,
in agreement with the results obtained in section~\ref{Eul-Brownian}.
However, as pointed out in \cite{Valageas2007a,Valageas2007b},
this decay does not express a true loss of memory associated with a relaxation
towards some equilibrium but it is only due to the random advection of the
density structures by the large-scale velocity effect (sweeping effect).
For the case of white-noise initial velocity the initial velocity variance is not well
defined but it becomes finite as soon as $t>0$ and the physics is the same.

Thus, the strong decay of the Eulerian propagator is due to the random advection
of the flow by the velocity field and it does not imply a true loss of memory
for the structures themselves. For instance, shifting the linear density field 
$\delta_L(x,t)$ by a random uniform translation $x \mapsto x+a L(t)$, with $a$
being distributed according to $P(U)$ of Eq.(\ref{RxJJ}), would lead 
to the same decaying propagator even though structures in the density field remain
unchanged.

\section{Lagrangian propagators}
\label{Lagrangian-propagators}

\subsection{Definitions and linear regime}
\label{Lag-general-expressions}

In order to go beyond the apparent loss of memory due to the ``sweeping effect''
discussed in the previous section, one method is to work within a Lagrangian
framework where the effect of uniform translations automatically vanishes.
Thus, following \cite{BernardeauVal2008}, we consider the Lagrangian 
propagator, $\Rkap(q,t;q_0)$, associated with the Lagrangian quantity
$\kappa(q,t)$ defined by
\beq
\kappa(q,t) = - \frac{\pl}{\pl q} [x(q,t)-q] = 1 - \frac{\pl x}{\pl q} ,
\;\;\;\; \mbox{whence} \;\;\; \kappa \leq 1 ,
\label{kappadef}
\eeq
where $x(q,t)$ is the Lagrangian map that describes the trajectory of particle $q$.
The upper bound, $\kappa \leq 1$, is associated with the fact that
particles do not cross each other, so that $x(q)$ is a monotonous
increasing function.
Thus, $-\kappa$ is the divergence of the Lagrangian displacement field 
$\chi(q,t)= x(q,t)-q$. It also describes the relative expansion of infinitesimal 
mass elements, and from Eq.(\ref{rhoxqvpsi}) it is related to the density field 
$\rho(x,t)$ as
\beq
\rho(x,t) = \frac{\rho_0}{1-\kappa(q,t)} , \;\;\; \mbox{whence} \;\;\;
\kappa(q,t) =  1 - \frac{\rho_0}{\rho(q,t)} , 
\label{rhokappa}
\eeq
with $\rho(q,t) \equiv \rho(x(q,t),t)$.
Note that if $x(q)$ is not monotonous (as would be the case for other systems
such as the collisionless gravitational dynamics) one would need to keep the 
absolute value, $|\pl q/\pl x|$, for the Jacobian that appears in the
expression of the density in (\ref{rhoxqvpsi}), which would violate the
relationship (\ref{rhokappa}) between $\kappa$ and $\rho$. In higher dimensions 
the relation (\ref{rhokappa}) no longer applies since the density is determined 
by the determinant of the deformation matrix, $|\pl \chi_i/\pl q_j|$, whereas $\kappa$
is defined as the trace of this matrix.
From a theoretical point of view, it is more convenient to work with $\kappa(q,t)$
than with the Lagrangian density $\rho(q,t)$, defined as $\rho(x(q,t),t)$,
because the Lagrangian equations of motion are usually most 
easily expressed in terms of the displacement field, $\chi(q,t)= x(q,t)-q$,
whence in terms of its divergence $-\kappa$ (e.g. \cite{BernardeauVal2008}).

As is clear from Eqs.(\ref{kappadef})-(\ref{rhokappa}), the variable $\kappa$
is not affected by uniform translations of the system, since it only depends on
the divergence of the displacement field. Therefore, it does not suffer from the
sweeping effect encountered for the Eulerian variables in 
section~\ref{Eulerian-propagators}, where we discussed Eulerian propagators.

In the linear regime, the Lagrangian map is simply given by $x=q+t u_0(q)$, 
and we have from (\ref{kappadef})
\beq
\kappa_L(q,t) = t \, \kappa_{L0}(q) \;\;\; \mbox{with} \;\;\; 
\kappa_{L0}(q) = - \frac{\dd u_0}{\dd q} = \frac{\dd^2 \psi_0}{\dd q^2} ,
\label{kappaL}
\eeq
whence
\beq
\kappat_{L0}(k) = - k^2 \psit_0(k) = \deltat_{L0}(k) ,
\label{kappaL0}
\eeq
where we used Eq.(\ref{rhopsi}).
Then, in a fashion similar to Eqs.(\ref{Rxdef}) and (\ref{Rrhodef}), we define the 
Lagrangian propagators $\Rkap(q,t;q_0)$ and $\Rkappsi(q,t;q_0)$ by
\beq
\Rkap(q,t;q_0) = \lag \frac{\cD\kappa(q,t)}{\cD\kappa_{L0}(q_0)} \rag ,
\;\;\; \Rkappsi(q,t;q_0) = \lag \frac{\cD\kappa(q,t)}{\cD\psi_0(q_0)} \rag .
\label{Rkapdef}
\eeq
For systems that are homogeneous and isotropic these propagators only depend
on the distance $|q-q_0|$ and on time, and going to Fourier space we obtain
\beq
\Rkappsit = -k^2 \Rkapt , \;\;\;\; 
\Rkappsi = \frac{\pl^2\Rkap}{\pl q^2}  =  \frac{\pl^2\Rkap}{\pl q_0^2} .
\label{Rkaptdef}
\eeq
For systems in the IR-class, $-3<n<-1$, without IR cutoff, we take the limit where
we are far from the reference point to obtain an homogeneous system.

In the linear regime, from Eqs.(\ref{kappaL})-(\ref{kappaL0}) the Lagrangian
propagators are given by
\beq
\Rkap_L = t \, \delta_D(q-q_0) , \;\;\; \Rkapt_L= t ,
\label{RkapL}
\eeq
\beq
\Rkappsi_L =  t \, \delta_D''(q-q_0) , \;\;\; \Rkappsit_L =  - t \, k^2 .
\label{RkappsiL}
\eeq
The asymptotic behaviors apply to the early-time ($t\rightarrow 0$) and
large-scale ($k\rightarrow 0$) limits for the whole range $-3<n<1$, as we
shall check below.

\subsection{Relation with the shock mass function}
\label{shock}

To compute the propagators beyond linear order, we first express $\kappa(q)$ in 
terms of the shocks built at a given time $t$, using the fact that all the matter
at any time $t>0$ is located within shocks for the fractional Brownian motion
initial conditions (\ref{ndef}) \cite{She1992}. Therefore, any Lagrangian point $q$
belongs almost surely to a shock at time $t>0$.
Thus, for a given realization 
of the initial velocity field, let us note $x_i$ the Eulerian position of the
shock $i$, which gathers the particles coming from the Lagrangian interval
$[q_i,q_{i+1}[$, with $x_i < x_{i+1}$ and $q_i < q_{i+1}$
(we may choose to affect the index $i=0$ to the first shock to the
right of $x=0$). In the case $-3<n<-1$, where shocks are dense, we only count
shocks above a finite mass $m_-$ to obtain a discrete sum and we eventually
take the limit $m_-\rightarrow 0$.
Then, for $q_i<q<q_{i+1}$ we have $x(q)=x_i$ whence $\kappa(q)=1$, whereas
the jump from $x_{i-1}$ to $x_i$ at point $q_i$ gives a contribution
$-(x_i-x_{i-1})\delta_D(q-q_i)$. Therefore, we can write $\kappa(q)$ as
\beq
\kappa(q) = 1 - \sum_{i=-\infty}^{\infty} (x_i-x_{i-1}) \, \delta_D(q-q_i) .
\label{kappashock1}
\eeq
Next, in order to perform the functional derivative (\ref{Rkapdef}) we must evaluate
the change of $\kappa(q)$ at linear order over a perturbation $\delta\psi_0(q_0)$.
For smooth initial potentials $\psi_0(q)$, the Lagrangian boundaries $\{q_i,q_{i+1}\}$
and the Eulerian location $x_i$ of the shock $i$ are obtained from the geometrical
construction (\ref{paraboladef}) which writes as the four constraints
\beq
\psi_0(q_i) = \cP_{x_i,c_i}(q_i)  \;\; \mbox{and} \;\; 
\psi_0'(q_i) = \cP_{x_i,c_i}'(q_i) ,
\label{psiPxc_i}
\eeq
and
\beq
\psi_0(q_{i+1}) = \cP_{x_i,c_i}(q_{i+1})  \;\; \mbox{and} \;\; 
\psi_0'(q_{i+1}) = \cP_{x_i,c_i}'(q_{i+1}) .
\label{psiPxc_i+1}
\eeq
They express the condition that the first-contact parabola $\cP_{x_i,c_i}$
simultaneously touches the curve $\psi_0$ at both points $\{q_i,q_{i+1}\}$,
with a tangent slope. Applying the perturbation $\delta\psi_0$ would give
at point $q_i$,
\beqa
\delta\psi_0(q_i) + \psi_0'(q_i) \delta q_i & = & \frac{q_i-x_i}{t} (\delta q_i - \delta x_i)
+ \delta c_i , \label{dpsi0} \\
\delta\psi_0'(q_i) + \psi_0''(q_i) \delta q_i & = & \frac{1}{t} (\delta q_i - \delta x_i) ,
\label{dpsi0p}
\eeqa
and similar relations at point $q_{i+1}$, from which we can derive the changes
$\delta q_i, \delta q_{i+1}, \delta x_i$, and $\delta c_i$, in terms of $\delta \psi_0$.
However, for the fractional Brownian motion initial conditions (\ref{ndef}) the changes
$\delta q_i$ and $\delta q_{i+1}$ vanish.

Let us first consider the IR-class, $-3<n<-1$. Then, from the scalings (\ref{scalingtheta0})
we can see that $\psi_0'=-u_0$ is finite but the second derivative $\psi_0''=\theta_0$
is almost surely infinite (i.e. $\psi_0$ has no regular second-order derivative; in the case
$n=-2$  for instance, $\psi_0''(q)$ is a white noise). More precisely, from 
(\ref{scalingtheta0}) we can see that the term $\psi_0''(q_i) \delta q_i$ in
Eq.(\ref{dpsi0p}), that measures the change $\Delta_i\psi_0'$ of $\psi_0'$ associated
with a shift $\delta q_i$, is of order $|\delta q_i|^{-(n+1)/2}$, so that $\delta q_i$ is of order
$|\delta \psi_0|^{-2/(n+1)}$ (e.g., for $n=-2$ where $\psi_0'(q)$ is a 
Brownian motion, $\Delta_i\psi_0' \sim \sqrt{|\delta q_i|}$ and
$\delta q_i \sim |\delta\psi_0|^2$). 
Therefore, the changes $\delta q_i$ and $\delta q_{i+1}$ are higher-order terms and
do not contribute at linear order to Eq.(\ref{dpsi0}).

On the other hand, for the UV-class, $-1<n<1$,  $\psi_0(q)$ has almost
surely no finite derivative and the change $\Delta_i\psi_0$ of $\psi_0$ associated
with a shift $\delta q_i$, is of order $|\delta q_i|^{(1-n)/2}$ (for instance,
$\Delta\psi_0 \sim \sqrt{|\delta q_i|}$ for the case $n=0$ of white-noise initial velocity).
Then, the tangent-slope constraints in Eqs.(\ref{psiPxc_i})-(\ref{psiPxc_i+1})
do not apply and the changes $\delta q_i$ and $\delta q_{i+1}$ are now non-perturbative,
so that there is no term $\psi_0'(q_i) \delta q_i$ in Eq.(\ref{dpsi0}).

Therefore, for the whole range $-3<n<1$, the Lagrangian boundaries $q_i$ of the
shocks do not change at linear order while the changes $\delta x_i$ and $\delta c_i$
are obtained from Eq.(\ref{dpsi0}), and its companion at $q_{i+1}$, setting 
$\delta q_i=\delta q_{i+1}=0$. This gives
\beq
\delta x_i = - t \, \frac{\delta\psi_0(q_{i+1}) - \delta\psi_0(q_i)}{q_{i+1}-q_i} .
\label{deltaxi}
\eeq
Substituting into Eq.(\ref{kappashock1}) yields
\beqa
\delta \kappa(q) & = & t  \sum_{i=-\infty}^{\infty} \delta_D(q-q_i) 
\left[ \frac{\delta\psi_0(q_{i+1}) - \delta\psi_0(q_i)}{q_{i+1}-q_i} \right. \nonumber \\
&& \left. - \frac{\delta\psi_0(q_i) - \delta\psi_0(q_{i-1})}{q_i-q_{i-1}} \right] ,
\eeqa
and the definition (\ref{Rkapdef}) gives
\beqa
\Rkappsi(q,t;q') & = &  t \; \lag \, \sum_i \delta_D(q-q_i) \times \nonumber \\
&& \hspace{-2.7cm}  \left[ \frac{\delta_{\!D}(q_{i+1}\!-\!q') \! - \! \delta_{\!D}(q_i\!-\!q')}
{q_{i+1}-q_i} - \frac{\delta_{\!D}(q_i\!-\!q') \! - \! \delta_{\!D}(q_{i-1}\!-\!q')}{q_i-q_{i-1}} 
\right] \rag . \nonumber \\
&&
\eeqa
This also reads as
\beqa
\hspace{-0.7cm} \Rkappsi(q,t;q') & = & t \; \lag \, \sum_i \delta_D(q\!-\!q_i) 
\nonumber \\
&& \hspace{-2.4cm} \times 
\frac{\delta_D(q\!-\!q'\!+\!\Delta_i q)+\delta_D(q\!-\!q'\!-\!\Delta_i q)-2\delta_D(q\!-\!q')}
{\Delta_i q} \, \rag 
\label{Rq2}
\eeqa
where we noted $\Delta_i q =q_{i+1}-q_i$ and we used the homogeneity and isotropy
of the system (so that $\Delta_{i-1} q$ and $\Delta_i q$ have the same statistical
properties).

Let us define the shock mass function at time $t$, $n(m,t) \dd m$, as the mean number
of shocks, per unit Eulerian or Lagrangian length (both functions are identical), with
a mass in the range $[m,m+\dd m]$. Since the initial density is uniform and equal
to $\rho_0$, the mass $m$ of a shock is related to its Lagrangian size $q$ (also
called the ``shock strength'') by $m=\rho_0 q$. Then, Eq.(\ref{Rq2}) writes as
\beqa
\hspace{-0.7cm} \Rkappsi(q,t;q_0) & = & t \int_0^{\infty} \dd m \, n(m,t) \nonumber \\ 
&& \hspace{-2.4cm} \times \frac{\delta_D(q\!-\!q_0\!+\!m/\!\rho_0)
+ \delta_D(q\!-\!q_0\!-\!m/\!\rho_0) - 2 \delta_D(q\!-\!q_0)}{m/\rho_0}
\label{Rkappsi_nm}
\eeqa
We can note from this expression that this Lagrangian propagator satisfies the
sum rule
\beq
\int \dd q  \, \Rkappsi(q,t;q_0) = \int \dd q_0  \, \Rkappsi(q,t;q_0) = 0.
\label{Rkappsi_sum}
\eeq
The propagator  $\Rkappsi(q,t;q_0)$ is singular at $q=q_0$ but for non-zero
separation Eq.(\ref{Rkappsi_nm}) simplifies as
\beq
q \neq q_0 : \;\; \Rkappsi(q,t;q_0) = t \rho_0 \frac{n(\rho_0|q-q_0|,t)}{|q-q_0|} .
\label{Rkappsi_nm1}
\eeq
We can check that expression (\ref{Rkappsi_nm}) only depends on the distance
$|q-q_0|$ and on time $t$, and it is clearly invariant through Galilean
transformations. As expected, we can see that the Lagrangian propagator $\Rkappsi$
provides a direct probe of the matter density field (which is directly related to the
structures of the displacement field, see Eqs.(\ref{rhoxqvpsi}) and (\ref{rhokappa}))
as it can be expressed in terms of the shock mass function.
Of course, the propagator $\Rkappsi$ alone is not sufficient to fully characterize the
density field (which would require for instance all $n$-point correlation functions)
but this represents a significant improvement over the Eulerian propagators
described in section~\ref{Eulerian-propagators}.
In the linear regime, where the typical shock mass goes to zero while the shock mass
function remains normalized to unity,
\beq
\int_0^{\infty} \dd m \, \frac{m}{\rho_0} \, n(m,t) = 1 ,
\label{norm_nm}
\eeq
Eq.(\ref{Rkappsi_nm}) clearly converges towards the first expression (\ref{RkappsiL}).
Going to Fourier space we obtain
\beq
\Rkappsit(k,t) = 2 t \int_0^{\infty} \dd m \, n(m,t) \, \frac{\cos(km/\rho_0)-1}{m/\rho_0} ,
\label{Rkappsit_nm}
\eeq
which agrees with the second expression (\ref{RkappsiL}) in the limits $t\rightarrow 0$
or $k\rightarrow 0$. Note that $\Rkappsi$ and $\Rkappsit$ are well defined over the
whole range $-3<n<1$, thanks to the property (\ref{norm_nm}) which ensures that there
is no divergence at low mass in Eq.(\ref{Rkappsit_nm}). Thus, contrary to the Eulerian
propagators studied in section~\ref{Eulerian-propagators}, there is no qualitative difference
between the UV-class and the IR-class for the Lagrangian propagator $\Rkappsi$
(however, for the IR-class where the initial velocity field only shows homogeneous
increments we must first take the limit of large distance from the reference point, to
avoid boundary effects).

\begin{figure}
\begin{center}
\epsfxsize=7 cm \epsfysize=5 cm {\epsfbox{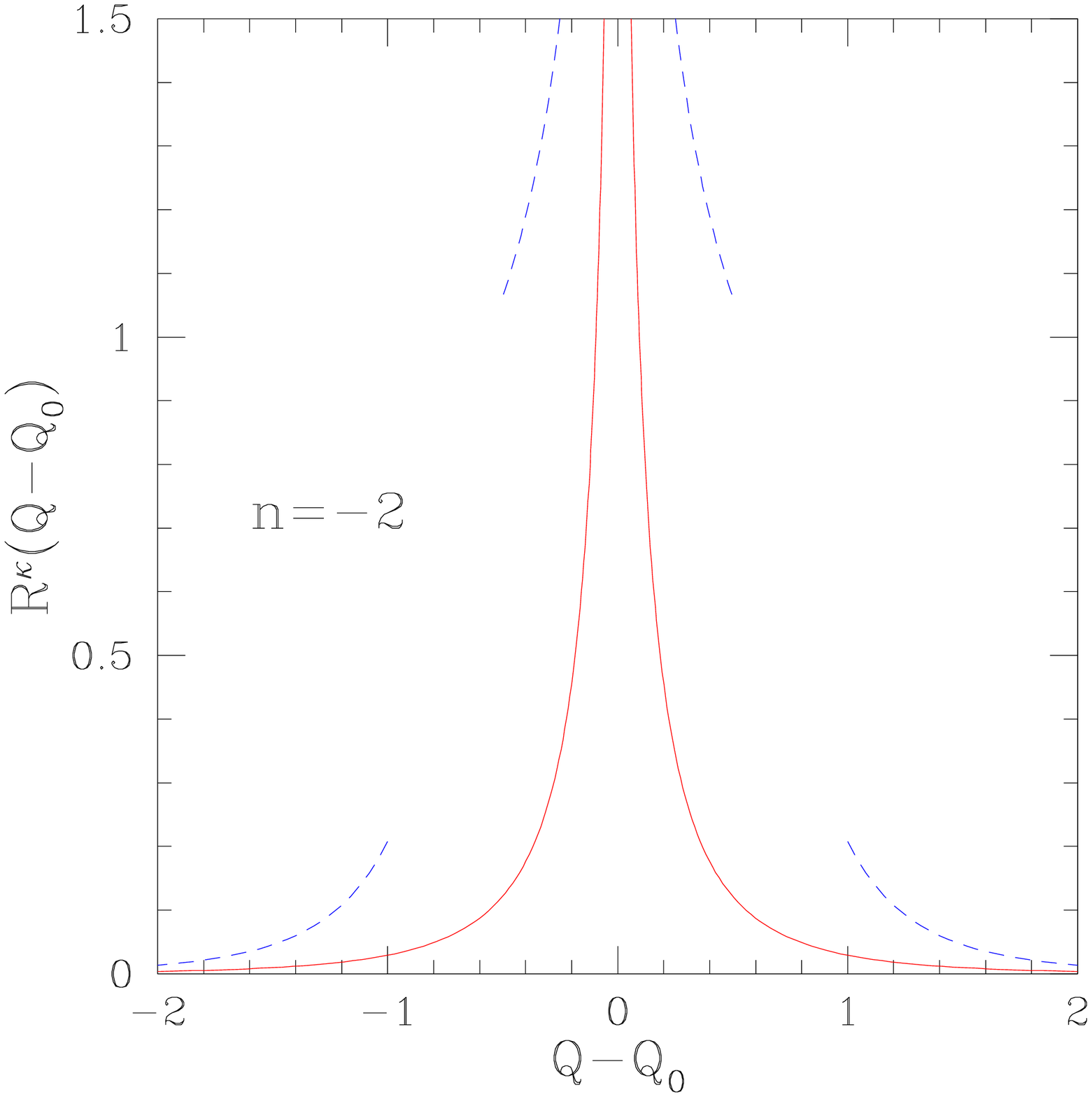}}
\epsfxsize=7 cm \epsfysize=5 cm {\epsfbox{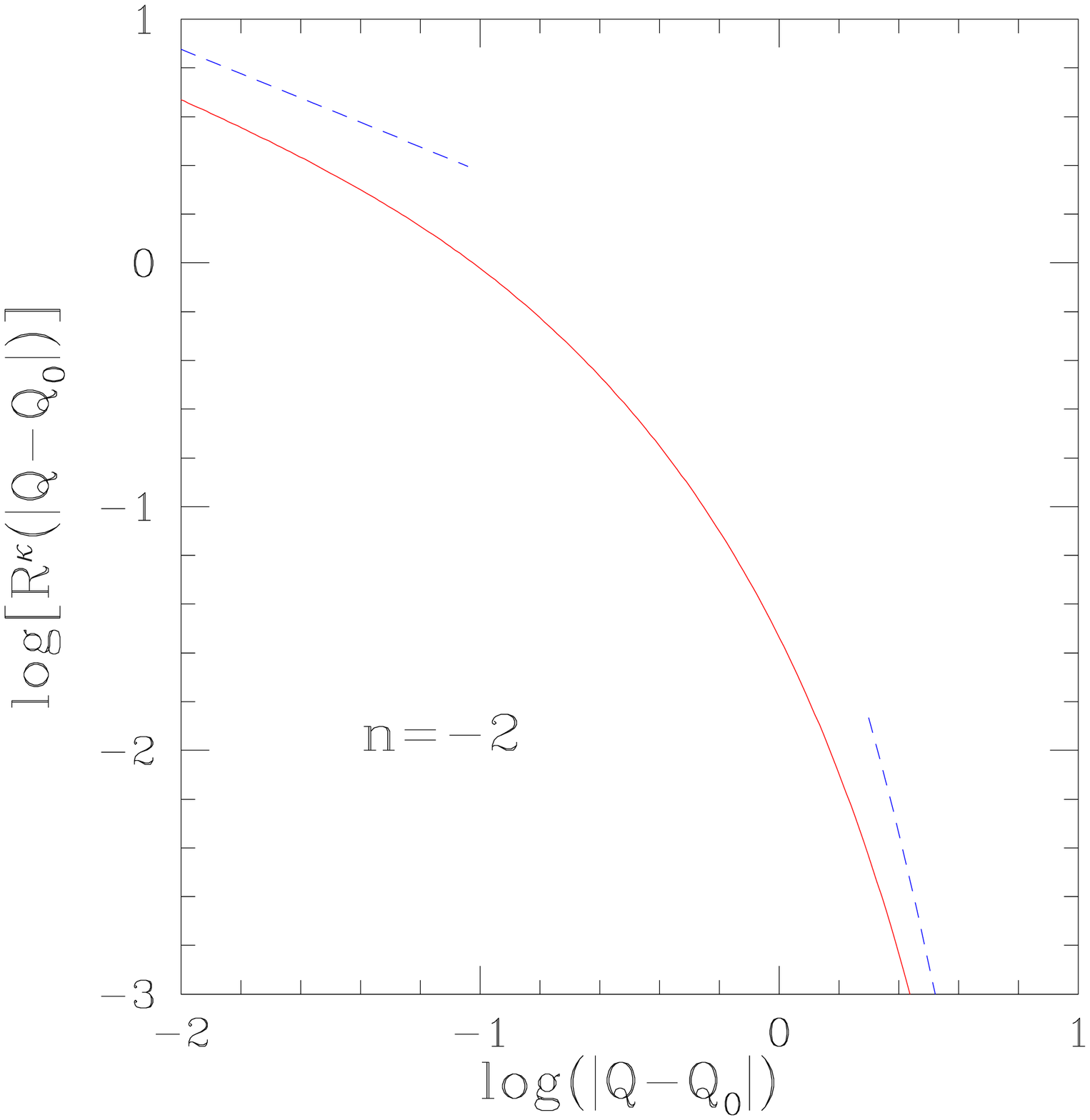}}
\end{center}
\caption{{\it Upper panel:} The Lagrangian propagator $\Rkap(Q-Q_0)$
obtained for $n=-2$, in terms of dimensionless variables, from
Eqs.(\ref{Rkap_nm1}) and (\ref{NM_n-2}).
The dashed lines are the asymptotic behaviors
(\ref{RkapQasympSmall_n-2})-(\ref{RkapQasympLarge_n-2}).
{\it Lower panel:} Same as upper panel but on a logarithmic scale.}
\label{figRkappaqBrown}
\end{figure}

\begin{figure}
\begin{center}
\epsfxsize=7 cm \epsfysize=5 cm {\epsfbox{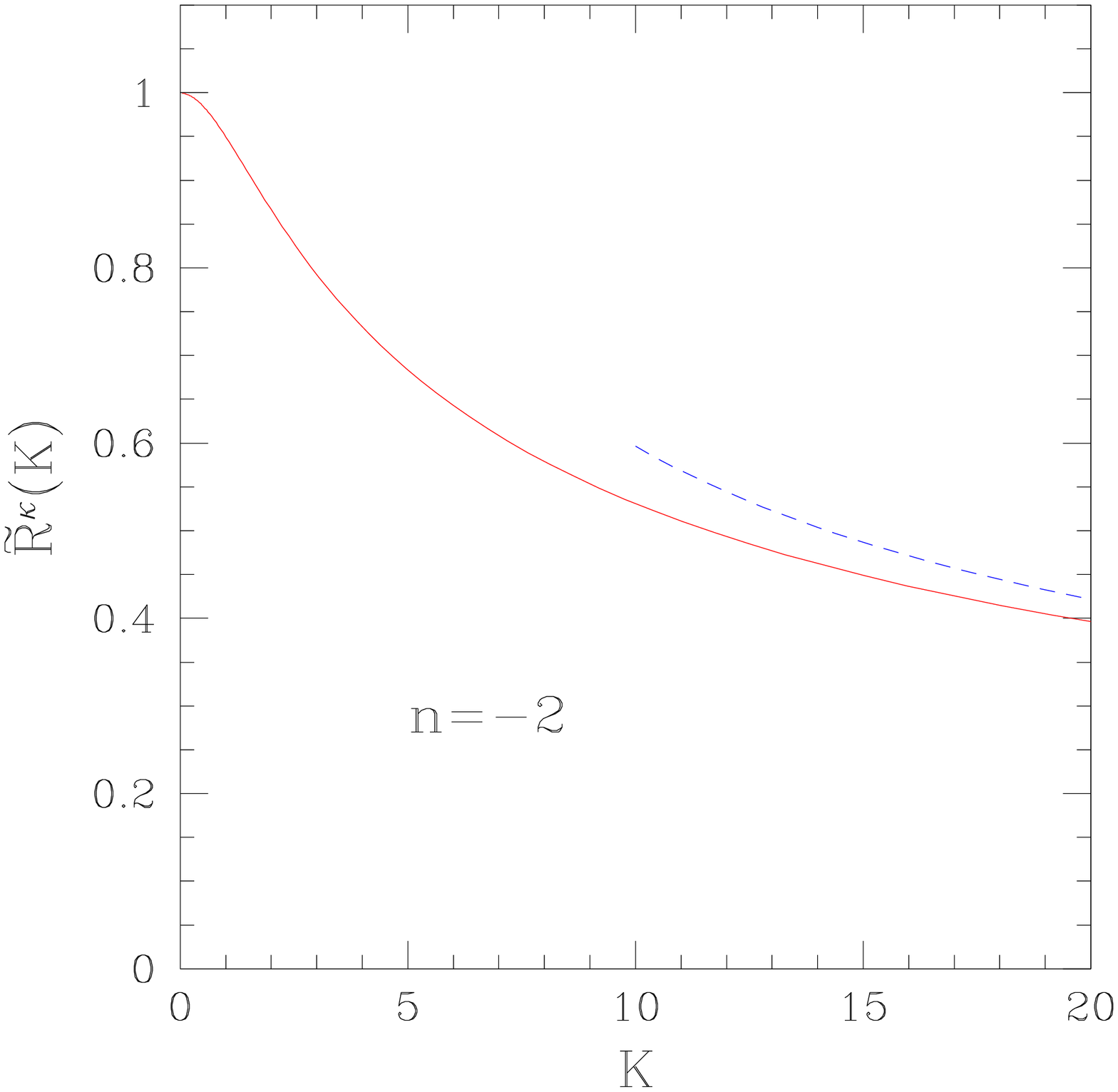}}
\epsfxsize=7 cm \epsfysize=5 cm {\epsfbox{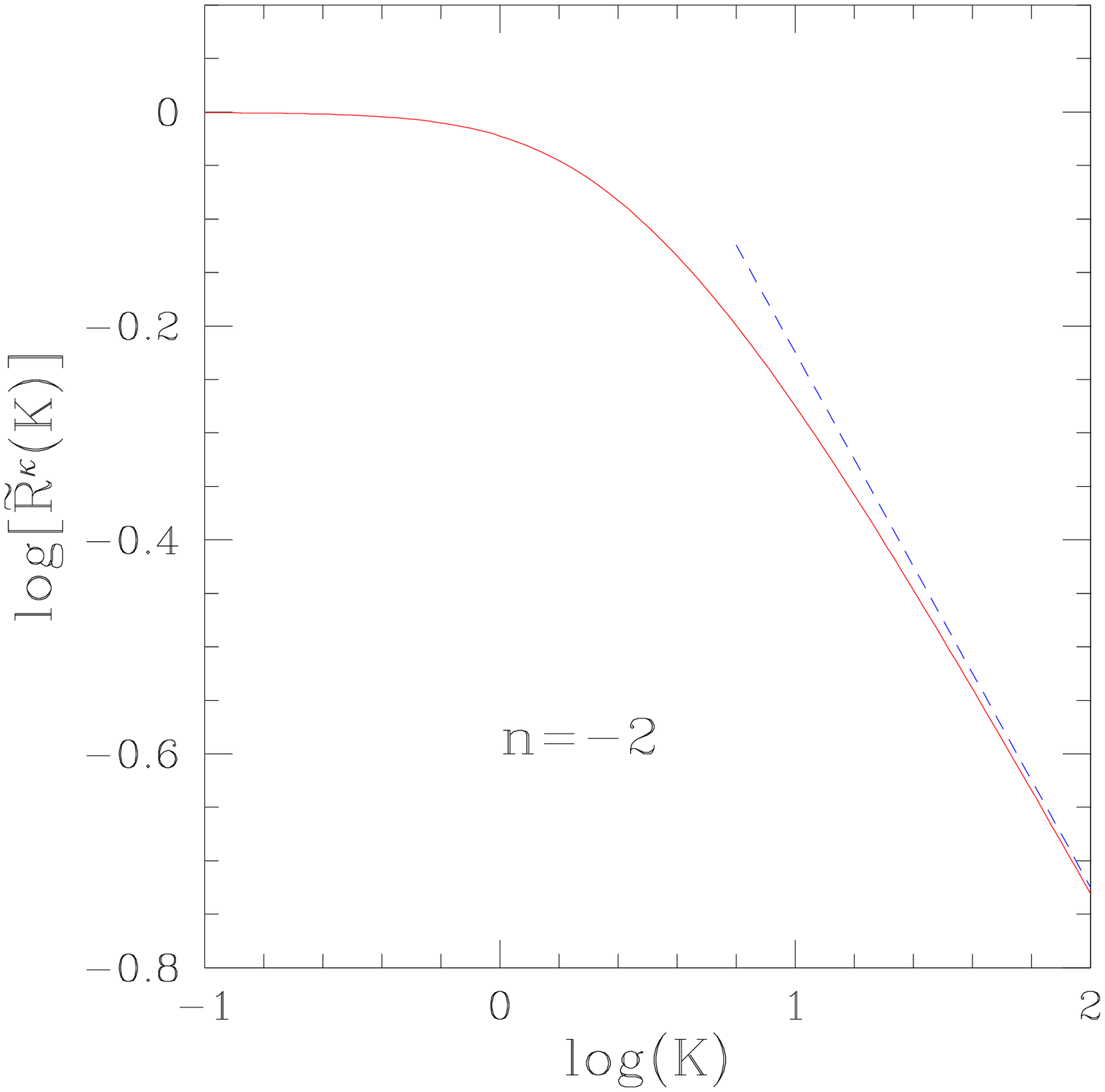}}
\end{center}
\caption{{\it Upper panel:} The Lagrangian propagator $\Rkapt(K)$
in Fourier space, obtained for $n=-2$, in terms of the dimensionless wavenumber
$K$, from Eqs.(\ref{Rkapt_nm}) and (\ref{NM_n-2}).
The dashed line is the asymptotic power-law behavior (\ref{RQKasymp_n-2}).
{\it Lower panel:} Same as upper panel but on a logarithmic scale.}
\label{figRkappakBrown}
\end{figure}

\begin{figure}
\begin{center}
\epsfxsize=7 cm \epsfysize=5 cm {\epsfbox{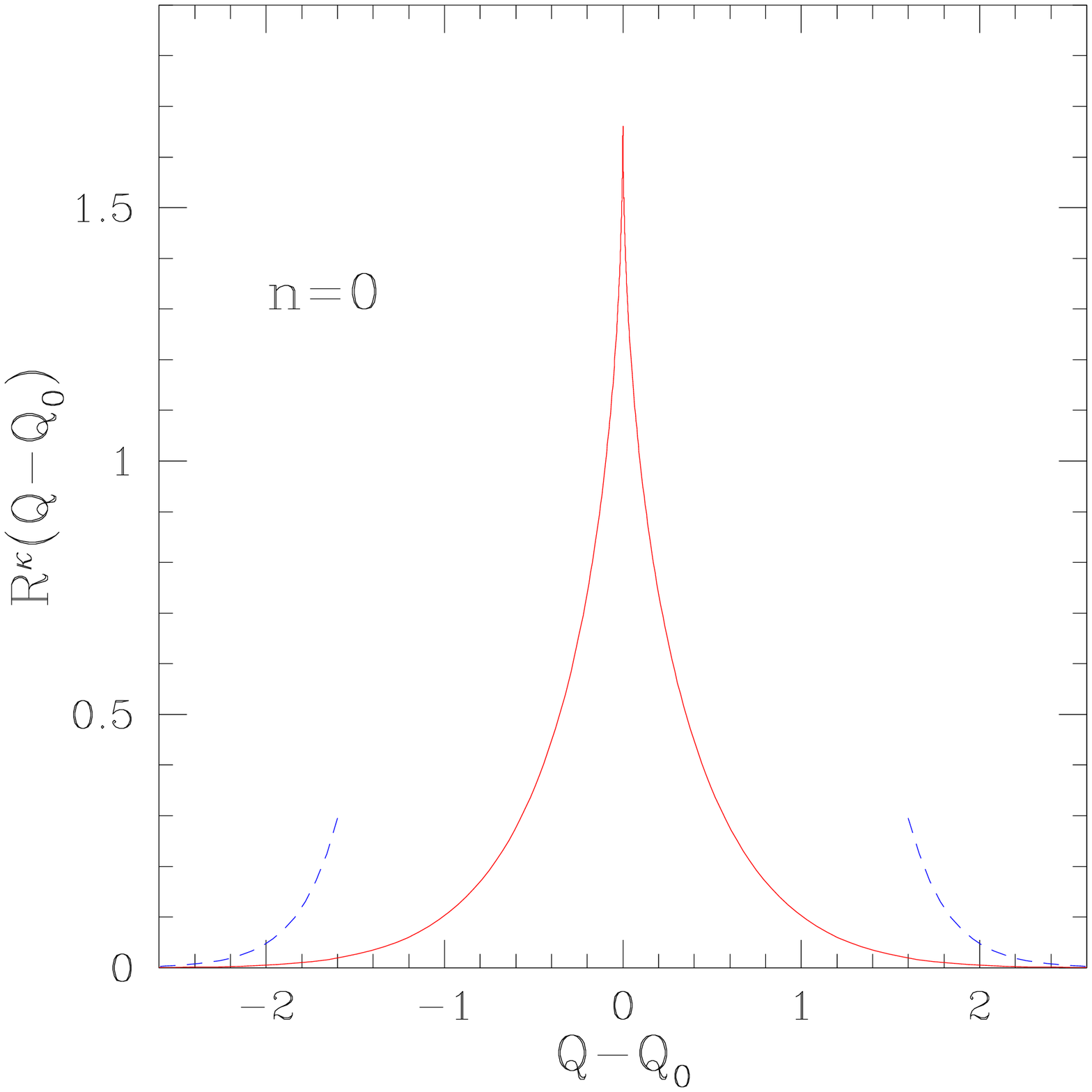}}
\epsfxsize=7 cm \epsfysize=5 cm {\epsfbox{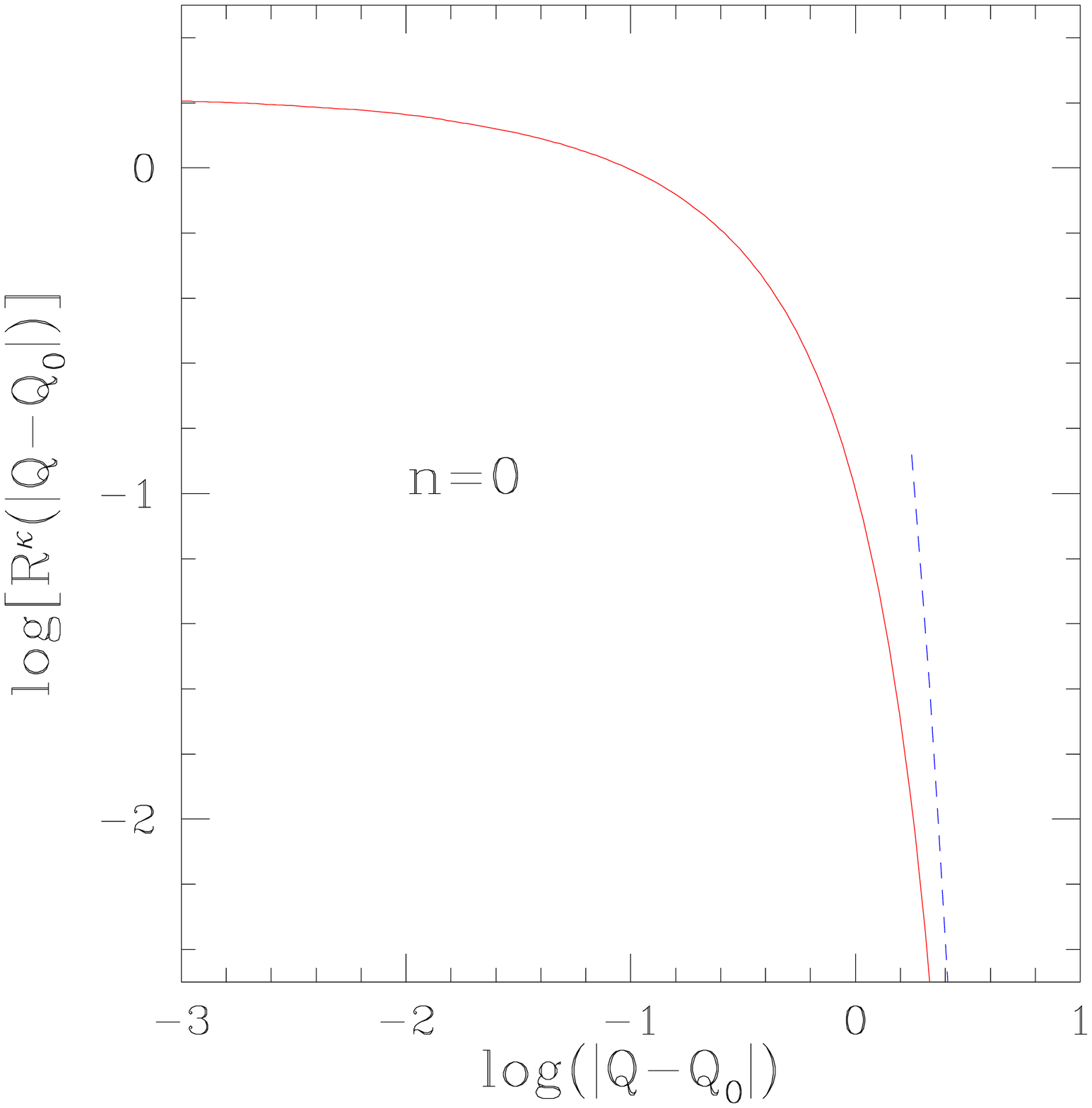}}
\end{center}
\caption{{\it Upper panel:} The Lagrangian propagator $\Rkap(Q-Q_0)$
obtained for $n=0$, in terms of dimensionless variables, from
Eqs.(\ref{Rkap_nm1}) and (\ref{NM_n0}).
The dashed lines are the asymptotic cubic exponential behavior
(\ref{RkapQasymp_n0}).
{\it Lower panel:} Same as upper panel but on a logarithmic scale.}
\label{figRkappaq}
\end{figure}

\begin{figure}
\begin{center}
\epsfxsize=7 cm \epsfysize=5 cm {\epsfbox{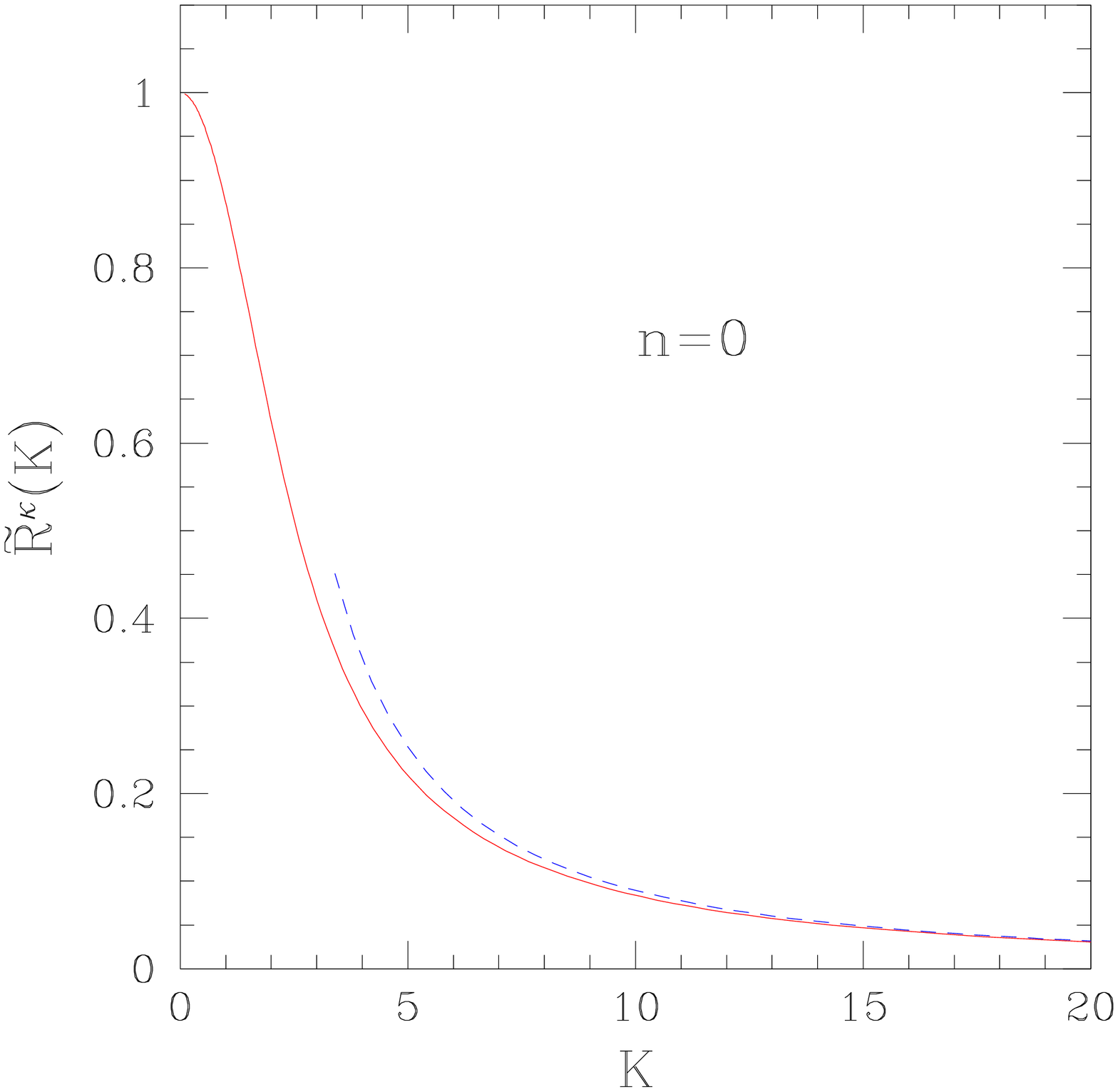}}
\epsfxsize=7 cm \epsfysize=5 cm {\epsfbox{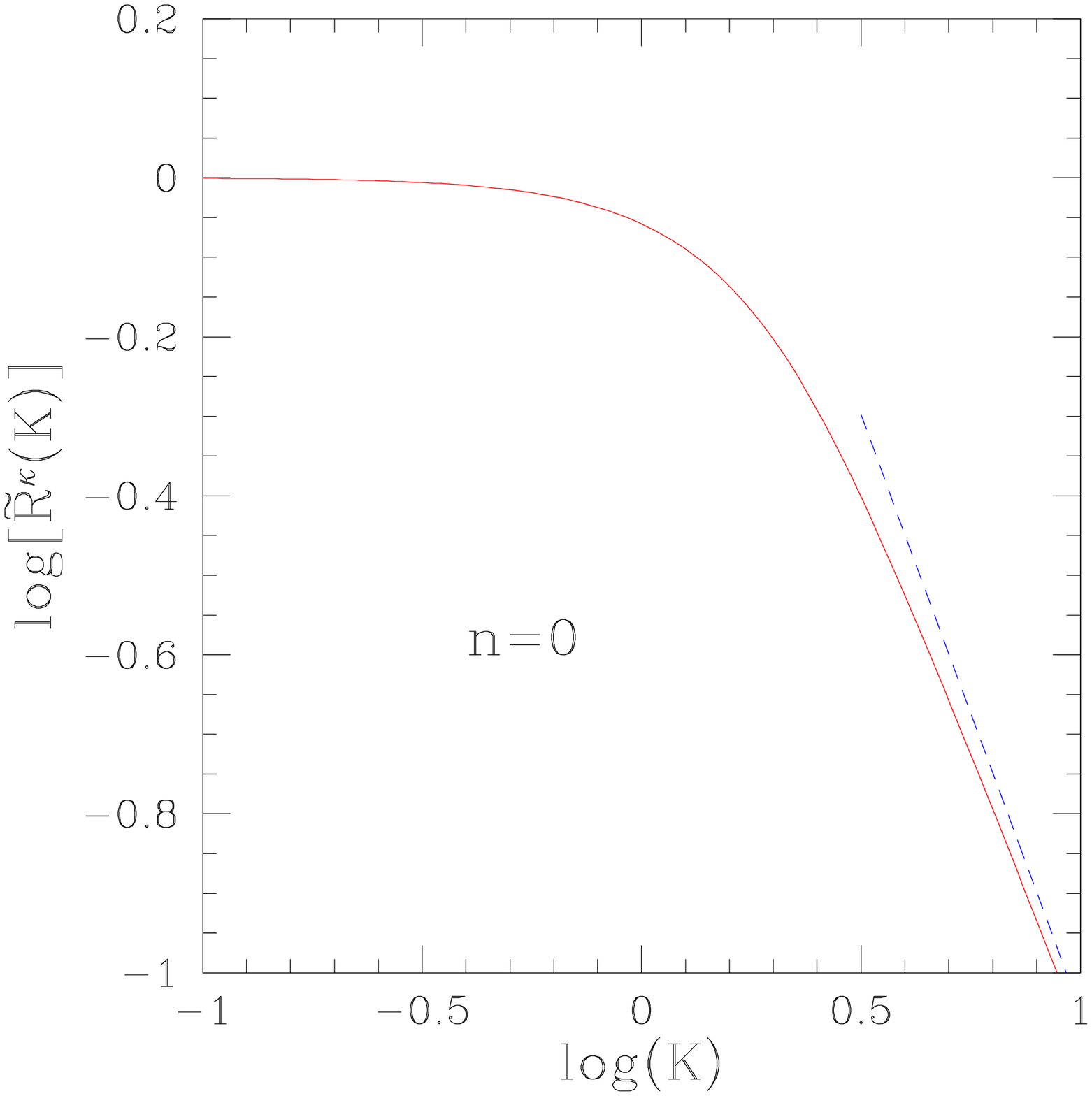}}
\end{center}
\caption{{\it Upper panel:} The Lagrangian propagator $\Rkapt(K)$
in Fourier space, obtained for $n=0$, in terms of the dimensionless wavenumber
$K$, from Eqs.(\ref{Rkapt_nm}) and (\ref{NM_n0}).
The dashed line is the asymptotic power-law behavior (\ref{RQKasymp_n0}).
{\it Lower panel:} Same as upper panel but on a logarithmic scale.}
\label{figRkappak}
\end{figure}

The Lagrangian propagator $\Rkap$ follows from Eqs.(\ref{Rkaptdef}), 
(\ref{Rkappsi_nm})  and (\ref{Rkappsit_nm}) as
\beqa
\hspace{-0.6cm} \Rkap(q,t;q_0) & = & t \int_0^{\infty} \dd m \, n(m,t) \nonumber \\ 
&& \hspace{-1.2cm} \times \frac{|q\!-\!q_0\!+\!m/\!\rho_0| + |q\!-\!q_0\!-\!m/\!\rho_0|
- 2 |q\!-\!q_0|}{2m/\rho_0} ,
\label{Rkap_nm}
\eeqa
which also reads as
\beq
\Rkap(q,t;q_0) =  t \int_{\rho_0|q-q_0|}^{\infty} \dd m \, n(m,t) \,
\left( 1 - \frac{|q-q_0|}{m/\rho_0} \right) ,
\label{Rkap_nm1}
\eeq
while we obtain in Fourier space
\beq
\Rkapt(k,t) = 2 t \int_0^{\infty} \dd m \, n(m,t) \, \frac{1-\cos(km/\rho_0)}{k^2m/\rho_0} .
\label{Rkapt_nm}
\eeq
Note that Eqs.(\ref{Rkap_nm})-(\ref{Rkap_nm1}) satisfy the constraint
$\Rkap(q,t;q_0) \rightarrow 0$ for $|q-q_0|\rightarrow \infty$, hence no additional
integration constant appears when we use Eq.(\ref{Rkaptdef}) to derive $\Rkap$
from $\Rkappsi$. 

As shown in \cite{Molchan1997,Ryan1998,Valageas2009b}, the shock mass function
shows the large-mass tail
\beq
-3 < n < 1 , \;\; m \gg \rho_0 L(t) : \;\; n(m,t) \sim e^{-m^{n+3}/t^2} ,
\label{nm_asympLarge}
\eeq
which is also related to the rare-event velocity tail (\ref{pu_asymp}). 
Again we did not write numerical factors in the exponential.
At low masses, numerical simulations and heuristic
arguments \cite{She1992,Vergassola1994} suggest the power-law tail
\beq
-3 < n < 1, \;\; m \ll \rho_0 L(t) : \;\; n(m,t) \sim t^{-1} \, m^{(n-1)/2} ,
\label{nm_asympSmall}
\eeq
which has only been proved rigorously for the Brownian case $n=-2$
\cite{Sinai1992,Bertoin1998,Valageas2009a} and for the white-noise
case $n=0$ \cite{AvellanedaE1995,Frachebourg2000}.
Then, Eqs.(\ref{nm_asympLarge}), (\ref{Rkappsi_nm1}), and (\ref{Rkaptdef}),
yield the large-separation asymptotic tails
\beq
\frac{|q-q_0|}{L(t)} \! \gg \! 1 \! :  \Rkappsi(q,t;q_0) \sim \Rkap(q,t;q_0)
\sim e^{-|q-q_0|^{n+3}/t^2}
\label{Rkap_Large}
\eeq
which show the same behavior as the tails of the Eulerian propagator
(\ref{Rpsi_asymp}). This is due to the fact that they are governed by the
same rare events, associated with extreme fluctuations in the initial
velocity field \cite{Valageas2009b}.
At small separations, Eqs.(\ref{nm_asympSmall}) and (\ref{Rkappsi_nm1}) give
\beqa
|q-q_0| \ll L(t) & : & \Rkappsi(q,t;q_0) \sim |q-q_0|^{(n-3)/2} , \;\;
\label{Rkappsi_Small} \\
&& \hspace{-2.2cm} \Rkap(q,t;q_0) \;\;  [ - \Rkap(0,t;0) ] \sim |q-q_0|^{(n+1)/2} , \;\;
\label{Rkap_Small}
\eeqa
with, for $-1<n<1$, 
\beq
\Rkap(0,t;0) = t \int_0^{\infty} \dd m \, n(m,t) \propto t^{(n+1)/(n+3)} .
\label{Rkap00t}
\eeq
The term $\Rkap(0,t;0)$ is only present in Eq.(\ref{Rkap_Small}) for
$-1<n<1$ (for $-3<n<-1$ the propagator $\Rkap(q,t;q_0)$ diverges for
$q\rightarrow q_0$ as $|q-q_0|^{(n+1)/2}$).
Note that the time-dependence (\ref{Rkap00t}) obtained for $-1<n<1$ is such
that we recover the linear behavior (\ref{RkapL}) on large scales, as we have
$\Rkap(0,t;0) L(t) \propto t$. For $-3<n<-1$ the linear time-dependence (\ref{RkapL})
is also recovered on large scales from Eq.(\ref{Rkap_Small}) through
$L(t)^{(n+1)/2+1} \propto t$.
On the other hand, for all $-3<n<1$, the time dependence vanishes in the leading-order
terms in the right hand side of Eqs.(\ref{Rkappsi_Small}) and (\ref{Rkap_Small}).

In Fourier space, from Eqs.(\ref{Rkappsit_nm}), (\ref{Rkapt_nm}),
we recover at low $k$ ($k\rightarrow 0$) the asymptotic limits
(\ref{RkapL}) and (\ref{RkappsiL}). At high $k$, the small-separation singularities
(\ref{Rkappsi_Small})-(\ref{Rkap_Small}) give rise to the time-independent
power-law tails
\beqa
k \gg L(t)^{-1} & : & \Rkappsit(k,t) \sim k^{(1-n)/2} , \label{Rkappsit_highk} \\
&& \Rkapt(k,t) \sim k^{-(n+3)/2} .
\label{Rkapt_highk}
\eeqa
Thus, $\Rkappsit(k,t)$ grows at high $k$ whereas $\Rkapt(k,t)$ slowly decreases,
in a time-independent fashion, over a range of wavenumbers that grows with
time since $L(t)^{-1}$ goes to zero at late times.

Note that, contrary to the Eulerian propagators discussed in section~\ref{Eul-Brownian},
the Lagrangian propagators are well defined over the whole range $-3<n<1$.
As explained in section~\ref{Lag-general-expressions}, the Lagrangian quantity
$\kappa(q,t)$ is not modified by uniform translations of the system, so that the
propagators $\Rkappsi$ and $\Rkap$ are not sensitive to long-wavelength modes
of the initial velocity and remain finite for $-3<n<-1$ as we push a possible IR cutoff
to infinity.

\subsection{Brownian and white-noise initial velocity}
\label{Brown-WN}

For the Brownian case, $n=-2$, the shock mass function is explicitly known
\cite{Bertoin1998,Valageas2009a},
\beq
n=-2 : \;\;\; N(M) = \frac{M^{-3/2}}{\sqrt{\pi}} e^{-M} ,
\label{NM_n-2}
\eeq
where we introduced the dimensionless scaling variables
\beq
M= \frac{m}{\rho_0 L(t)} , \;\;\; n(m,t) = \frac{1}{\rho_0 L(t)^2} N(M) .
\label{Mdef}
\eeq
Introducing the dimensionless propagators,
\beq
\Rkap(q,t;q_0) = \frac{t}{L(t)} \Rkap(Q-Q_0) , \;\;\;
\Rkapt(k,t)= t \, \Rkapt(K) ,
\label{Rlag_nodim}
\eeq
we obtain from Eq.(\ref{Rkappsi_nm1})
\beq
\frac{\dd^2\Rkap}{\dd Q^2}(Q) = \frac{N(Q)}{Q} =  \frac{Q^{-5/2}}{\sqrt{\pi}} e^{-Q} ,
\eeq
whence
\beqa
Q \ll 1 & : & \;\; \Rkap(Q) \sim \frac{4}{3\sqrt{\pi Q}} , \label{RkapQasympSmall_n-2} \\
Q \gg 1& : & \;\; \Rkap(Q) \sim \frac{Q^{-5/2}}{\sqrt{\pi}} e^{-Q} ,
\label{RkapQasympLarge_n-2}
\eeqa
as well as $\Rkapt(0)=1$ and
\beq
K \gg 1 : \;\; \Rkapt(K) \sim \frac{4}{3} \sqrt{\frac{2}{K}} .
\label{RQKasymp_n-2}
\eeq

For the white-noise case, $n=0$, the shock mass function is again explicitly known
\cite{Frachebourg2000,Valageas2009c},
\beqa
n=0 : \;\;\; N(M) & \!\! = \!\! & 2 M \,  \inta\frac{\dd s'}{2\pi\ii} \, 
\frac{e^{-s' M}}{\Ai(s')^2} \nonumber \\
&& \times  \inta\frac{\dd s}{2\pi\ii} \, e^{s M} \, \frac{\Aip(s)}{\Ai(s)}  ,
\label{NM_n0}
\eeqa
using the dimensionless scaling variables (\ref{Mdef}).
It shows the asymptotic tails \cite{Frachebourg2000}
\beqa
M \gg 1 & : & N(M) \sim  2\sqrt{\pi} \, M^{5/2} \, e^{-\om_1 M - M^3/12} , \;\; \\
M \ll 1 & : & N(M) \sim \frac{1}{\sqrt{\pi M}} , 
\eeqa
which agree with Eqs.(\ref{nm_asympLarge})-(\ref{nm_asympSmall}).
In terms of the dimensionless propagators (\ref{Rlag_nodim}) we obtain
$\Rkap(0) \simeq 1.674$ and the asymptotic tail
\beq
Q \gg 1 : \;\; \Rkap(Q) \sim 32 \sqrt{\pi} \, Q^{-5/2} \, e^{-\om_1 Q-Q^3/12} , 
\;\;\;\;
\label{RkapQasymp_n0}
\eeq
as well as $\Rkapt(0)=1$ and
\beq
K \gg 1 : \;\; \Rkapt(K) \sim 2 \sqrt{\frac{2}{K^3}} .
\label{RQKasymp_n0}
\eeq

We show in Figs.~\ref{figRkappaqBrown}-\ref{figRkappak} the propagators
$\Rkap(Q-Q_0)$ and $\Rkapt(K)$ obtained for the two cases $n=-2$ and $n=0$.

\subsection{Evolution with time}
\label{Evolution-with-time}

\begin{figure}
\begin{center}
\epsfxsize=7 cm \epsfysize=5 cm {\epsfbox{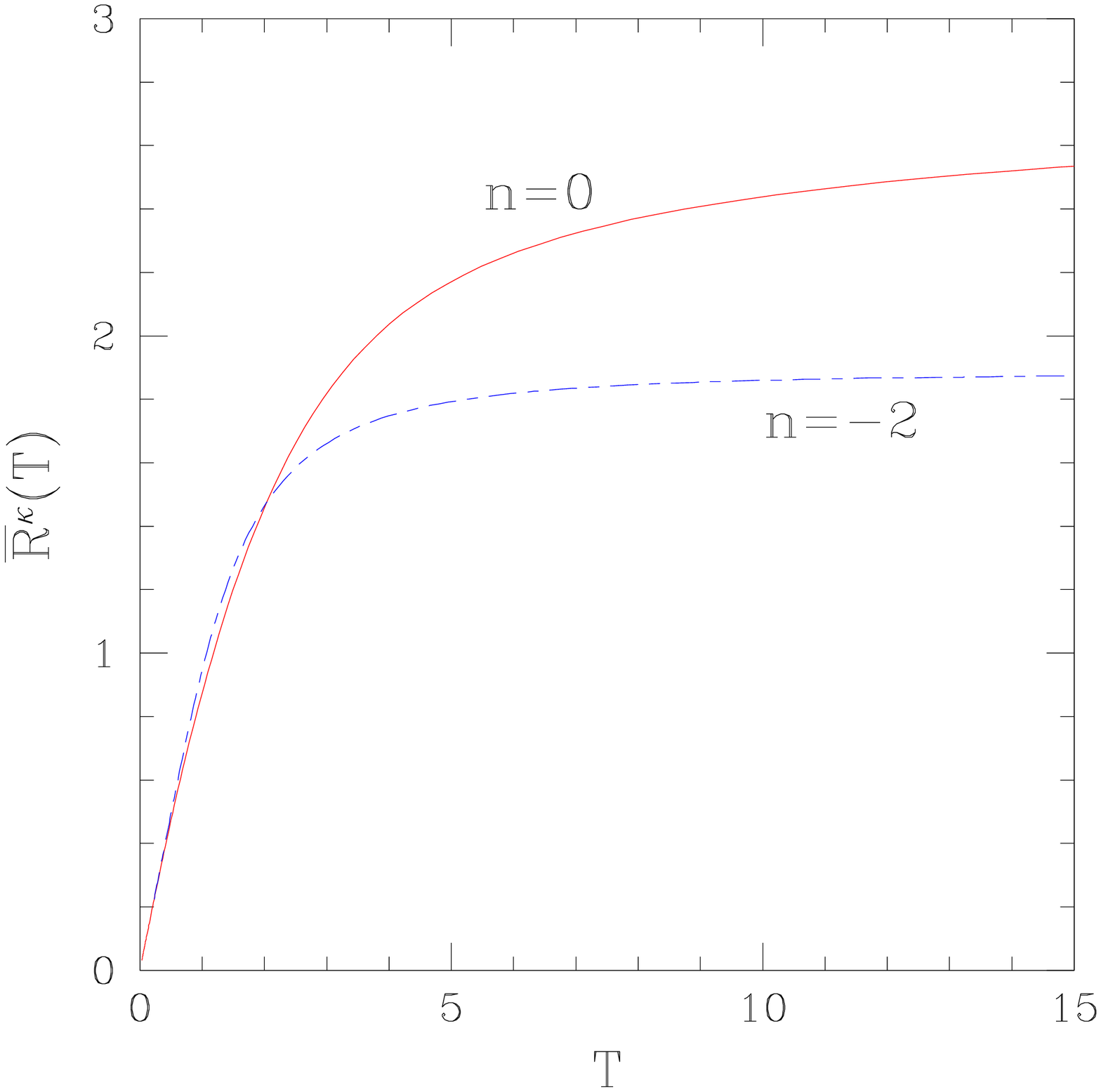}}
\epsfxsize=7 cm \epsfysize=5 cm {\epsfbox{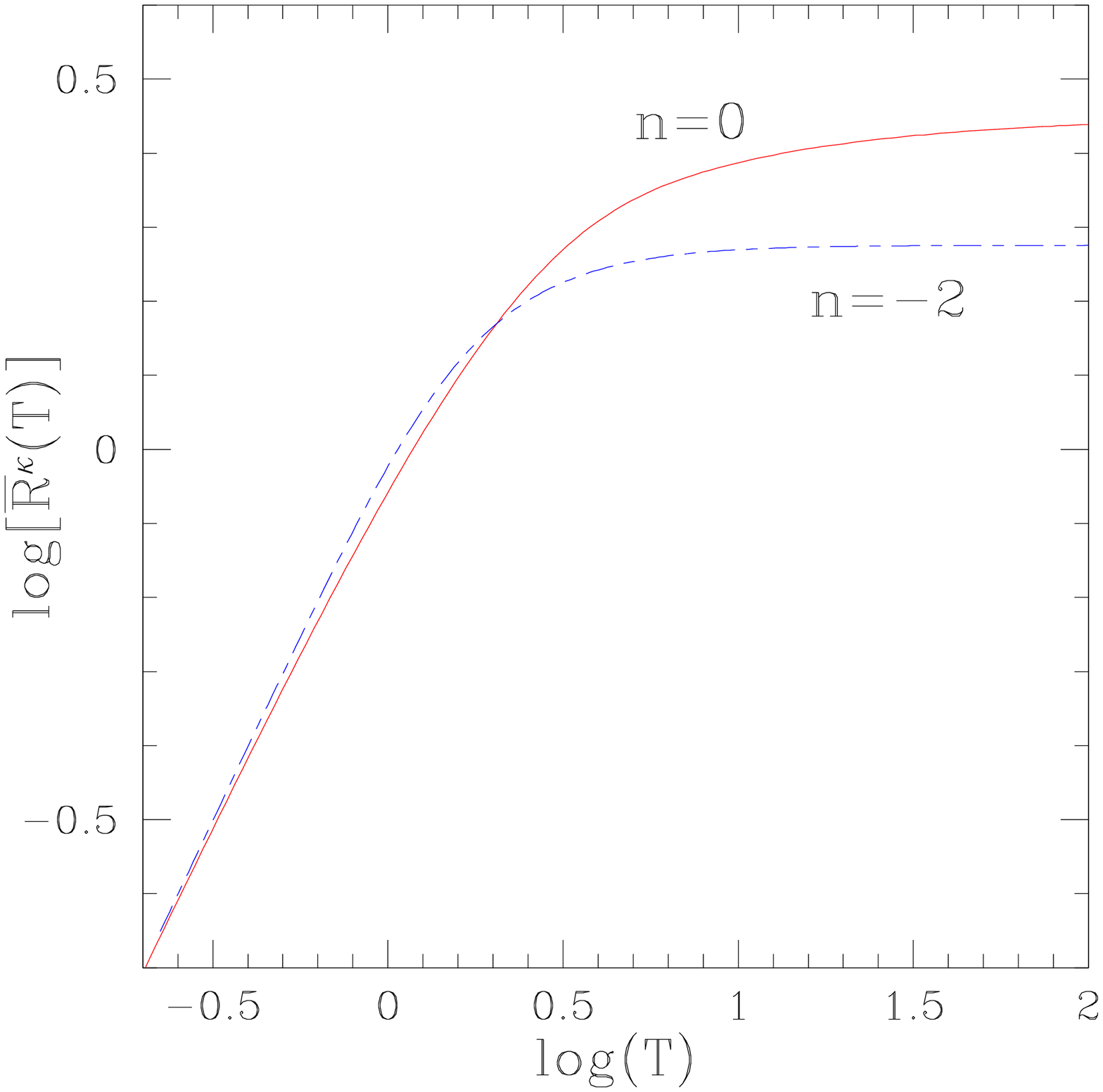}}
\end{center}
\caption{{\it Upper panel:} The Lagrangian propagator $\Rkapb(T)$
as a function of the dimensionless time $T=t/t_k$, at fixed wavenumber $k$.
The dot-dashed line corresponds to Brownian initial velocity ($n=-2$) and the
solid line to white-noise initial velocity ($n=0$).
{\it Lower panel:} Same as upper panel but on a logarithmic scale.}
\label{figRkapb}
\end{figure}

As shown in the previous section, for a fixed wavenumber $k$ we recover 
at early times the linear growth (\ref{RkapL})-(\ref{RkappsiL}), whereas at late times
the Lagrangian propagators saturate into the nonlinear regime, as seen in
Eqs.(\ref{Rkappsit_highk})-(\ref{Rkapt_highk}).
To emphasize the evolution with time, we define the characteristic time $t_k$,
associated with wavenumber $k$, which measures the time when scales $\sim 1/k$
enter the nonlinear regime (i.e. $K=1$), by
\beq
k L(t_k) = 1 , \;\;\; \mbox{whence} \;\;\; t_k \propto k^{-(n+3)/2} ,
\label{tkdef}
\eeq
where $L(t)$ was introduced in Eq.(\ref{Lt}), and we introduce the associated
dimensionless time-variable $T$,
\beq
T = \frac{t}{t_k} , \;\;\; \mbox{whence} \;\;\; T = K^{(n+3)/2} .
\label{TKdef}
\eeq
Then, we write the propagator $\Rkapt(k,t)$ of Eq.(\ref{Rlag_nodim}) as
\beq
\Rkapt(k,t) = t_k \, \Rkapb(T) \;\; \mbox{whence} \;\; \Rkapt(K) = \frac{1}{T} \Rkapb(T) .
\label{Rkapbdef}
\eeq
As noticed above, the propagator $\Rkapb(T)$, which describes the time-dependence
of $\Rkapt(k,t)$ at fixed $k$, shows the early- and late-time asymptotics
\beq
T \ll 1 : \Rkapb(T) \sim T , \;\;\;\; 
T \rightarrow \infty : \;\; \Rkapb(T) \rightarrow \mbox{constant} .
\label{Rkapbasymp}
\eeq
We obtain $\Rkapb(\infty)=4\sqrt{2}/3$ for $n=-2$ and $\Rkapb(\infty)=2\sqrt{2}$
for $n=0$.
We display in Fig.~\ref{figRkapb} the scaling propagator $\Rkapb(T)$ obtained
for $n=-2$ and $n=0$.

The behaviors (\ref{Rkapbasymp}) explicitly show that the Lagrangian propagators
first grow linearly with time until they reach the nonlinear regime where they saturate,
by contrast with the Eulerian propagators, which show exponential-type
decays as seen in section~\ref{Eulerian-propagators}.
The fact that the former show no decay at late times implies that there is
no real loss of memory.

\section{Conclusion}
\label{Conclusion}

In this article we have studied the Eulerian and Lagrangian propagators which are
obtained within the adhesion model (i.e. Burgers dynamics), focusing on the
one-dimensional case with power-law initial conditions that develop a self-similar
evolution. We have derived some general and asymptotic results for linear
power-spectrum index $n$ in the range $-3<n<1$, as well as complete explicit  
expressions for the two representative cases $n=-2$ (Brownian initial velocity)
and $n=0$ (white-noise initial velocity). In particular, we note that the range
$-3<n<1$ can be split over an ``IR class'', $-3<n<-1$, governed by long wavelengths,
and a ``UV class'', $-1<n<1$, governed by small wavelengths.
 
We find that Eulerian propagators can be expressed in terms of the one-point
Eulerian velocity probability distribution. This clearly shows that they are sensitive
to long-wavelength modes of the velocity field, which leads to a ``sweeping effect''
as small-scale structures can be moved over large distances without significant
distortions if most of the power is stored at very low wavenumbers. This yields
a strong exponential-like decay at high-$k$ of Eulerian propagators, which does not
imply a strong loss of memory of the system as it is due to this random advection.
In the IR-class, the Eulerian propagators show an universal Gaussian decay
(i.e. independently of $n$) of the form $e^{-t^2k^2\sigma^2_{u_0}/2}$ in Fourier space.
However, if there is no infrared cutoff the factor $\sigma^2_{u_0}$ diverges 
and the Eulerian propagators vanish as soon as $t>0$. 
In the UV-class, Eulerian propagators strongly depend on the index $n$.
In particular, for $n=0$ we obtain in Fourier space an oscillatory decay, with an amplitude
of the form $e^{-t k^{3/2}}$. 
In both IR and UV classes, being related to the one-point Eulerian velocity probability
distribution Eulerian propagators are not sensitive probes of the structures of the
underlying density field.

For more complex dynamics, such as cosmological gravitational clustering, Eulerian
propagators can no longer be written in terms of the velocity distribution in such
a direct manner, but they remain governed by the same sweeping effect and
resummation schemes typically show a decay at high wavenumbers
and late times \cite{Crocce2006a,Crocce2006b,Valageas2007a,BernardeauVal2008}.
Even when this effect is modest (i.e. there is not much power at low $k$) 
Eulerian propagators are unlikely to provide good probes of the density field, its relation
to the one-point velocity distribution being rather loose.

Next, we have shown that Lagrangian propagators can be expressed in terms of the
shock mass function (which corresponds to the halo mass function in the cosmological
context) as shown explicitly in Eqs.~(\ref{Rkappsi_nm}), (\ref{Rkappsit_nm}) and
(\ref{Rkapt_nm}). 
Therefore, Lagrangian propagators  are much more closely related to the properties of the
density field. Moreover, they show the same properties for both the IR and UV classes,
as could be expected from the Galilean invariance of the equations of motion.
Whereas on large scales and low $k$ they grow with time, in agreement with the linear
regime, on small scales and high $k$ they saturate to constant values, with a power-law
dependence on scale or wavenumber that is set by the initial index $n$.
This strong memory of the system marks a sharp contrast with the decay of
different-time Eulerian statistics.
In higher dimensions we still expect the Lagrangian propagators to show power-law
tails over wavenumber $k$ in the highly nonlinear regime, with a slope that is again
related to the low-mass tail of the shock mass function, but there can also be
a power-law decay over time. We leave such a study to future work.

Even though in more complex dynamics Lagrangian propagators
are unlikely to be expressed in terms of the density field in such a direct manner
(i.e. through the mass function of bound objects), they should provide a sensitive
probe of the behavior of the density field and of the relaxation processes at work.
As seen in this article, a great interest of the adhesion model is to provide exact
results in a nontrivial case which shares some key properties with cosmological
gravitational clustering. This allows us to obtain some insight in the behavior of
complex quantities, such as propagators or response functions, and the processes
which they probe. Then, this can serve as a guide to decipher the processes associated
with more complex systems.
On a quantitative level, it remains to be seen whether this can be used to build
for instance efficient ansatze (in the manner of the halo model \cite{2002PhR...372....1C} or the
stable-clustering ansatz \cite{Peebles1980}) that would help building an accurate
description of the system.

\begin{acknowledgments}
This work is supported in part by the French Agence Nationale de la Recherche under grant ANR-07-BLAN-0132 
(BLAN07-1-212615).
\end{acknowledgments}

\bibliography{ref-burgers1D}   

\end{document}